\documentclass[a4paper,fleqn,usenatbib,useAMS]{mnras}

\usepackage{graphicx}	% Including figure files
\usepackage{amsmath}	% Advanced maths commands
\usepackage{amssymb}	% Extra maths symbols
\usepackage{multicol}        % Multi-column entries in tables
\usepackage{bm}		% Bold maths symbols, including upright Greek
\usepackage{pdflscape}	% Landscape pages
\usepackage{color} 

\usepackage[T1]{fontenc}
\usepackage{ae,aecompl}

\title[Multi-phase outflows in IRAS17]{NOEMA spatially resolved view of the multi-phase outflow in IRAS17020+4544: a shocked wind in action?}

\author[A.L. Longinotti et al.]{Anna Lia Longinotti$^{1}$\thanks{Contact e-mail: \href{mailto:annalia@inaoep.mx}{alonginotti@astro.unam.mx}}, Quentin Salom\'e$^{2,9}$, Chiara Feruglio$^{3}$, Yair Krongold$^{1}$, \newauthor Santiago Garc\'ia-Burillo$^{4}$,  Marcello Giroletti$^{6}$, Francesca Panessa$^{7}$, Carlo Stanghellini$^{6}$, \newauthor Olga Vega$^{5}$, Victor Manuel Pati\~no-\'Alvarez$^{5,11}$, Vahram Chavushyan$^{5,10}$,  \newauthor Mauricio El\'ias-Chavez$^{5}$, Aitor Robleto-Or\'us$^{1,8}$ \\
$^{1}$Instituto de Astronom\'ia, Universidad Nacional Aut\'onoma de M\'exico,  Circuito Exterior, Ciudad Universitaria, Ciudad de M\'exico 04510, M\'exico\\
$^{2}$ Finnish Centre for Astronomy with ESO (FINCA), University of Turku, Vesilinnantie 5, 20014 Turku, Finland \\
$^{3}$INAF-Osservatorio Astronomico di Trieste, via G. Tiepolo 11, I-34143 Trieste, Italy \\
$^{4}$ Observatorio Astron\'omico Nacional (OAN-IGN)-Observatorio de Madrid, Alfonso XII, 3, 28014-Madrid, Spain\\
$^{5}$ Instituto Nacional de Astrof\'isica, \'Optica y Electr\'onica, Luis E. Erro 1, Tonantzintla, Puebla, M\'exico, C.P. 72840\\ 
$^{6}$ INAF Istituto di Radioastronomia, via Gobetti 101, 40129, Bologna Italy\\
$^{7}$Istituto di Astrofisica e Planetologia Spaziali di Roma (IAPS), Via del Fosso del Cavaliere 100, 00133 Roma, Italy\\
$^{8}$ Departamento de Astronom\'ia, Universidad de Guanajuato, Apdo. 144, C.P. 36000 Guanajuato, Gto., Mexico;\\
$^{9}$ Aalto University Mets\"ahovi Radio Observatory, Mets\"ahovintie 114, 02540 Kylm\"al\"a, Finland\\
$^{10}$ Center for Astrophysics  Harvard \& Smithsonian, 60 Garden Street, Cambridge, MA 02138, USA\\
$^{11}$ Max-Planck-Institut f\"ur Radioastronomie, Auf dem H\"ugel 69, D-53121 Bonn, Germany\\
}

\pubyear{2023}

\begin{document}
\label{firstpage}
\pagerange{\pageref{firstpage}--\pageref{lastpage}}
\maketitle

% Abstract of the paper
\begin{abstract}
The Narrow Line Seyfert 1 Galaxy IRAS17020+4544 is one of the few AGN where  a galaxy-scale energy-conserving outflow was revealed.
This paper reports on NOEMA observations addressed to constrain the spatial scale of the CO emission in outflow.
The molecular outflowing gas is  resolved in five components tracing approaching and receding gas, all located at a distance of 2-3~kpc on the West  and East side of the active nucleus. This high velocity gas (up to v$_{out}$=$\pm$1900~km~s$^{-1}$) is  not coincident with the  rotation pattern of the CO gas in the host galaxy disk.  The estimated mass outflow rate shows that with a global mass output of $\dot{M}_{H_2}$=~139$\pm$20$~M_\odot$~yr$^{-1}$,  this powerful galaxy-scale outflow  is consistent with the wind conserving its energy, and with a momentum rate boost of a factor of $\sim$30 compared to the momentum rate of the nuclear X-ray wind. 
  Preliminary results from ancillary X-ray ({\it Chandra}) and radio images (e-MERLIN) are reported. While the nature of the radio source is not conclusive, the  {\it Chandra} image  may  tentatively trace extended emission, as expected by an expanding bubble of hot X-ray gas. The outcome of the  NOEMA analysis and of past and ongoing publications dedicated to the description of the outflow multi-band phenomenology  in IRAS17020+4544 concur to provide compelling reasons to postulate that an outflow shocking with the galaxy interstellar medium  is driving the multi-phase wind in this peculiar AGN.
  \end{abstract}

% Select between one and six entries from the list of approved keywords.
% Don't make up new ones.

%%%%%%%%%%%%%%%%%%%%%%%%%%%%%%%%%%%%%%%%%%%%%%%%%%

%%%%%%%%%%%%%%%%% BODY OF PAPER %%%%%%%%%%%%%%%%%%

% The MNRAS class isn't designed to include a table of contents, but for this document one is useful.
% I therefore have to do some kludging to make it work without masses of blank space.
%\begingroup
%\let\clearpage\relax
%\tableofcontents
%\endgroup
%\newpage

\section{Introduction}
The well-established relations observed between the properties of galaxies and their nuclear black hole activity (Kormendy \& Ho 2013)  have suggested the presence of a solid yet still mysterious mechanism that allows the black hole behaviour at nuclear scale to impact the environment at galaxy-scale (Silk \& Rees 1998, King 2003). This proposition finds strong support in a myriad of theoretical models and hydrodinamical simulations of galaxy formation and evolution developed through the last decades (e.g. Di Matteo et al. 2005, Hopkins \& Elvis 2010, Weinberger et al. 2017, Pillepich et al. 2018) that concur to indicate feedback by AGN as a key ingredient for regulating  star formation and clearing up of the gas in galaxies (Zubovas \& King 2012).  

In this regard, large outflows  of gas in galaxies have long been seen as a viable agent for feedback due to their action of carrying mass and energy from the innermost regions out to larger scales  via ejection of jets and winds driven by the AGN, as currently observed in AGN data collected over several bands of the electromagnetic spectrum (see review by Harrison et al. 2018).
Most of the numerous models proposed to explain the role of AGN driven outflows in feedback processes, postulate that this complex mechanism can be started  by a relativistic, fast wind launched close to the AGN  accretion disk  (see review by King \& Pounds 2015 and reference therein).  Over the last decade, X-ray observations of Ultra Fast Outflows (UFO) in Seyfert Galaxies  have provided evidence for the existence of  such relativistic nuclear winds moving at {\it v}~$\ge$0.1{\it c} in around 40\% of the studied AGN samples (Tombesi et al. 2010 with 42 objects;  Gofford et al. 2013 with 51 objects). 

As a result of the impact of this high velocity wind with the galaxy interstellar medium, it is expected that an expanding  two-phase shock is generated where dense cold clumps of molecular material co-exist with gas at much higher temperature (e.g. Faucher-Giguere \& Quataert 2012, Zubovas \& King 2014). 
Depending on the cooling properties of this expanding shocked outflow, specifically if the expansion is adiabatic, the initial energy of the nuclear wind can be efficiently transported  outward and ``transferred" to the kpc-scale outflows of molecular gas that are indeed frequently observed in luminous AGN (e.g. Veilleux et al. 2013, Cicone et al. 2014, Aalto et al. 2015, Alonso-Herrero et al. 2018, Garcia-Burillo et al. 2019). 

Under the assumption of energy-conservation, the wind  momentum rate at large scale undergoes a boost that is proportional to the  ratio of the wind outflow velocities at nuclear and large scale (Tombesi et al. 2015, Feruglio et al. 2015). In this regard, to a first approximation,  comparing the energetics of the X-ray and molecular outflow phases  provides a test for the presence of energy-conserving outflows, and, ultimately, a probe of how efficient is the coupling between the UFO and the host galaxy interstellar medium.  From an observational perspective (e.g. Cicone et al. 2018), grasping the specifics of this mechanism is extremely challenging since feedback processes encompass the entire galaxy activity, therefore multi-band observations are needed to cover outflow properties spanning from the active nucleus (X-ray winds) to the galaxy outskirts (atomic, molecular and ionized gas outflows).

Observations of luminous AGN where this comparison was available, show that few sources follow the energy-conserving scenario. 
For instance, from the compilation of the 10 sources from the literature where properties of nuclear and galaxy-scale winds  are well-established, Marasco et al. (2020) concluded that less than half of them are consistent with energy conservation, which may suggest some inefficiency of the mechanism responsible for mass and energy transportation. 

The radio-loud Seyfert Galaxy  IRAS17020+4544 belongs to this compilation: past observations of a sub-relativistic X-ray wind with {\it XMM-Newton} (Longinotti et al. 2015) and of a powerful molecular outflow with the Large Millimeter Telescope  have shown that the energy is efficiently transported from the nuclear region to galaxy scale thanks to a boost in the outflow momentum of a factor of $\sim$60 (Longinotti et al. 2018).

The main goal of the present paper is to report on the spatial distribution of the outflowing molecular gas previously observed with the LMT single-dish antenna by exploiting the higher resolution of the observations carried out with the NOEMA interferometer. These data are complemented by radio (e-MERLIN) and  X-rays ({\it Chandra}) observations aimed to characterize possibly extended emission in these bands. Because the main focus of this publication is the spatially resolved study of the molecular outflow, the technical description of the  X-ray (\ref{subsec:xray}) and radio (\ref{subsec:radio}) ancillary  data are presented in the Appendix. Only their outcome and preliminary results are discussed in the main body of the paper. 

 Section~\ref{sec:observations} describes previous multi-band observations of IRAS17020+4544 (\ref{sec:iras17}), and the NOEMA (\ref{subsec:NOEMA}), Chandra (\ref{subsec:Chandra}) and e-MERLIN (\ref{subsec:merlin}) data. 
The NOEMA data analysis and its results are respectively reported in Section~\ref{mol_out} and Section~\ref{sec:results}.
The Discussion is presented in Section~\ref{sec:discussion}.  
Section~\ref{sec:summary} summarizes the main findings of this paper.

\section{Observations: the past and the present}
\label{sec:observations}
\subsection{The observational history of the AGN IRAS17020+4544}
\label{sec:iras17}
IRAS17020+4544 (IRAS17 hereafter) is  classified as a Narrow Line Seyfert 1 and as a Luminous Infrared Galaxy in IR band (LIRG). The first indication of the radio-loud nature of this source came from Gu \& Chen (2010), who reported a milli-arcsecond jetted structure at 5 GHz.
Later on, a  VLBA study by  Doi et al. (2011), confirmed the presence of a one-sided jet at 1.7 GHz, with a projected extension of about 35~pc.  In their VLBI study,  Giroletti et al. (2017) revealed that the synchrotron radiation observed in these data at 5 GHz came from a compact yet elongated structure on a scale of $\sim$10~pc. Recent reports on JVLA data by J{\"a}rvel{\"a} et al. (2022) have shown that star formation processes measured by mid-infrared colours are not sufficient to account for the observed luminosity.  All these studies concurred to say that the radio  emission in IRAS17 is produced in a jet, yet not well-collimated and observed under a large viewing angle.
  
   As to X-rays, IRAS17 was one of the first AGNs  where the detection of a  multi-component X-ray UFO in {\it XMM-Newton} grating spectra (Longinotti+2015) showed a stratified structure far more complex than previously envisaged for fast X-ray outflows (e.g. Tombesi et al. 2010).  Under the conservative  assumption of a bolometric luminosity of L$_{bol}$$\sim$ 5$\times$10$^{44}$ ergs s$^{-1}$ (see Longinotti et al. 2015 for details),  the most massive of the five components detected in X-rays was estimated to expel sufficient mass for inducing feedback in the host galaxy.  
 Subsequently, the findings of a composite system of slower X-ray absorbers, also confirmed by {\it Chandra} spectroscopy (Longinotti 2020), has led Sanfrutos et al. (2018) to propose that a shocked outflow may explain the intricate X-ray and radio properties of this source.  
 
More recently, the first study of the ultraviolet spectrum of this source provided by the {\it HST-COS} spectrograph (Mehdipour et al. 2022) has unveiled the rare presence (e.g. Mehdipour et al. 2023) of a fast UV outflow detected in Ly$\alpha$ absorption at 23,400 km~s$^{-1}$. This wind has been identified as the ultraviolet counterpart of one of the X-ray UFO low ionization components, pointing to a scenario where the UV gas revealed in absorption by {\it HST-COS} is formed by entrainment and shock of the X-ray UFO with the surrounding interstellar medium. 

As can be seen, the previous results on the outflowing gas in IRAS17 listed above are based mostly on spectroscopic information. 
Optical imaging of the host galaxy of this interesting AGN by Ohta et al. (2007) indicated the presence of a central bar in the galaxy although more recent data obtained with the Nordic Optical Telescope did not confirm complex morphology (Olgu\'in-Iglesias et al. 2020).

The first spatially resolved study of the properties of the molecular gas in the host galaxy reported in our companion paper (Salom\'e et al. 2021) and based on the same NOEMA dataset herein presented,  has revealed several unexpected facts about IRAS17. 
 While the bulk of the CO gas ($\sim$10$^9$~M$_\odot$) is distributed in a central disc-like structure of about 4~kpc of radius, a high concentration of molecular gas ($\sim$10$^8$~M$_\odot$) located up to 
 $\sim$8~kpc north of the nucleus characterized by  different dynamics with respect to IRAS17, points to the previously unknown presence of a satellite galaxy. This discovery provides evidence for some interaction of IRAS17 with  a low mass system  (probably in the first stages of a minor merger), very much at odds  with the former vision of it as the typical undisturbed spiral galaxy.
 
 The analysis of the NOEMA CO spectrum has highlighted that the double-peaked line emitted by the molecular gas (Fig.6 in Salom\'e et al. 2021) presents a significant displacement from the position expected by  the optical redshift ({\it z$_{opt}$}=0.0604, de Grijp et al. 1992),corresponding to $\sim$225 km~s$^{-1}$. This shift was also noted in the CO line observed by the LMT telescope (see Fig.5 in Salom\'e et al. 2021) and the difference with the value derived from the optical emission lines was attributed to the possible effect of outflowing gas in the Narrow Line Region.   The new CO-estimated redshift of  {\it z$_{CO}$}=0.0612 is then adopted throughout this paper.  Finally, in their analysis of the gas content of the whole galaxy,  Salom\'e et al. 2021 also reported the presence of a CO outflow of about 10$^7$~M$_\odot$ moving at a velocity of $\sim$780~km~s$^{-1}$, located in projection of the northern companion galaxy and therefore unlikely to be related to the current AGN activity. A brief discussion on this component, herein called ``Northern Outflow" or ``Outflow N"  is included in \ref{subsec:outflow_N}.

\subsection{NOEMA interferometry}
\label{subsec:NOEMA}

IRAS17 was observed in CO(1-0) at a redshifted frequency of 108.623 GHz with NOEMA in May 2018 under the project number W17CR. The data were obtained with 9 antennas in C configuration  (baselines from 15 to 290m, which makes these observations sensitive to a spatial scale of 45$^{\prime\prime}$) for a total observing time of 4.8h on-source. Data were imaged and cleaned using the GILDAS package {\tt mapping}. Further technical details of these observations and their data reduction are described in the companion paper (Salom\'e et al. 2021). In that work, after subtracting the continuum in the {\tt uv} plane, the spectral resolution was smoothed to 4~MHz $\sim$ 11\: km~s$^{-1}$ reaching an rms of 1~mJy/beam. However, we note that the CO outflow observed with LMT presented a full-width at half maximum (FWHM) greater than 1000~km~s$^{-1}$ (Longinotti et al. 2018). Therefore,  here we decided to  smooth the spectral resolution of the NOEMA data to  20~MHz~$\sim$ 55~km~s$^{-1}$ to reach a rms of 0.5 mJy/beam and improve the signal-to-noise ratio.
 The spatial resolution of the two data cubes in both works (the present and Salom\'e's et al. 2021) corresponds to a  synthesised beam of 2.2$^{\prime\prime}$$\times$1.7$^{\prime\prime}$ (PA$\sim$ 43$^\circ$).
 The NOEMA data analysis is described  in Section~\ref{mol_out}.

\subsection{{\it Chandra} X-ray imaging}
\label{subsec:Chandra}
{\it Chandra} observed IRAS17 in 9 visits distributed from November 2016 to February 2017 and for a total exposure of 250~ks (Sequence Number 703217). 
The main aim of this campaign was to obtain {\it LETG} (Low Energy Transmission Grating) spectroscopy of the X-ray outflow to complement the {\it XMM-Newton}  spectroscopic study fully reported in Longinotti et al. 2015 and Sanfrutos et al. 2018. While preliminary results of the Chandra LETG spectroscopy  can be found in Longinotti (2020), the entire study of this high-resolution spectrum will be presented in a dedicated paper (Longinotti et al. in prep.).  

However, in this paper, we decided to include imaging data obtained by  the High Resolution Camera (HRC-S), which was operated as readout detector of the LETG during the {\it Chandra} campaign. The technical details of the data reduction and analysis of the {\it Chandra} image of IRAS17 are reported in  the Appendix (Section \ref{subsec:xray}). 

Here we anticipate that the HRC detector, with a nominal spectral coverage of 0.1-10 keV, unfortunately does not allow precise energy separation of the incident photons\footnote{https://cxc.cfa.harvard.edu/ciao/threads/hrc\_note/}. In practical terms, this means that the resulting HRC image cannot provide information as to which specific energy band the X-ray photons are distributed in. Moreover, the instrumental configuration with the LETG grating in place for the whole campaign, makes the spatial analysis of the HRC X-ray image far from standard (see Section \ref{subsec:xray}). 
Despite the limited amount of spatial and spectral information that could be extracted from this imaging data, its inclusion in this manuscript (see Figure~\ref{fig:Chandra}) provides a broader context to the discussion on the galaxy-scale outflowing gas.

 \subsection{Radio observations: e-MERLIN}
 \label{subsec:merlin}
 Prompted by the intriguing results on the radio properties of this source reported by Giroletti et al. (2017), in the recent years our group has dedicated significant effort to gain insights on the radio emission of IRAS17, which has been the target of several radio campaigns including VLBA, e-MERLIN, VLA. Given the extent of their output, the study of the radio properties using data from these campaigns will be reported in a separate publication (Stanghellini  et al. in prep.). 
 
 However, motivated by the same aim to include the X-ray image, we decided to hereby report a preliminary outcome of the e-MERLIN campaign, which obtained  observations at 1.51 GHz on 14 February and 12/13 March 2020.  As for the Chandra data, the technical details on the e-MERLIN observations are described in Section~\ref{subsec:radio}. We include the view of the radio source at the corresponding spatial scale of these observations  ($\sim$ 1arcsec) in  Figure~\ref{fig:Chandra}.

\section{NOEMA data analysis}
\label{mol_out}

\begin{figure*}
  \centering
  \includegraphics[height=7cm,trim=0 0 85 0,clip=true]{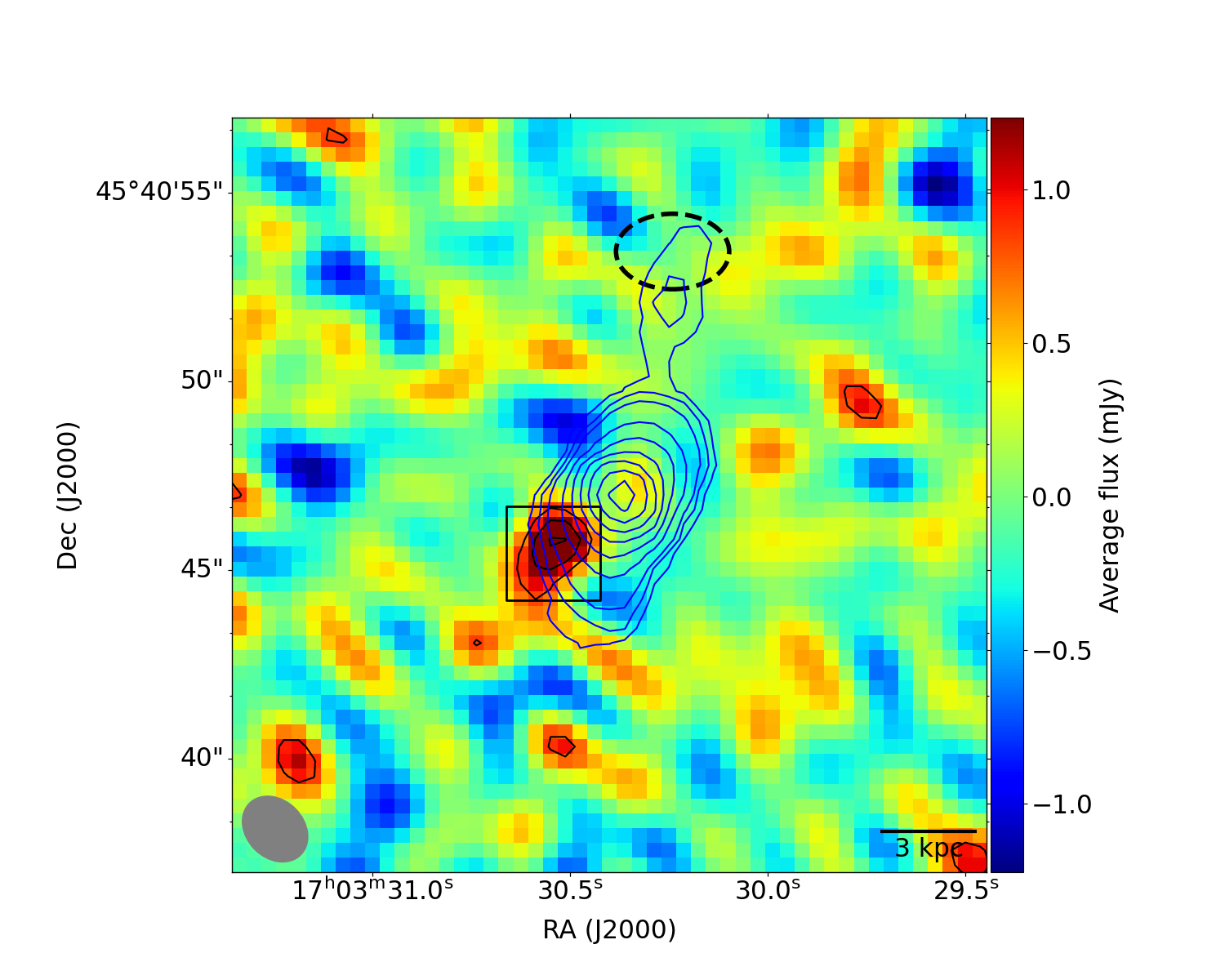}
  \hspace{3mm}
  \includegraphics[height=6cm,trim=15 25 50 50,clip=true]{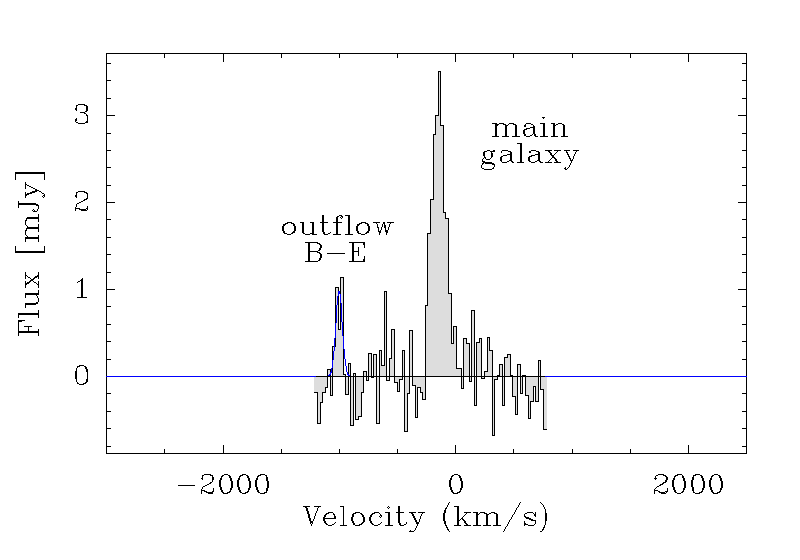} \\
  \vspace{3mm}
  \includegraphics[height=7cm,trim=0 0 85 0,clip=true]{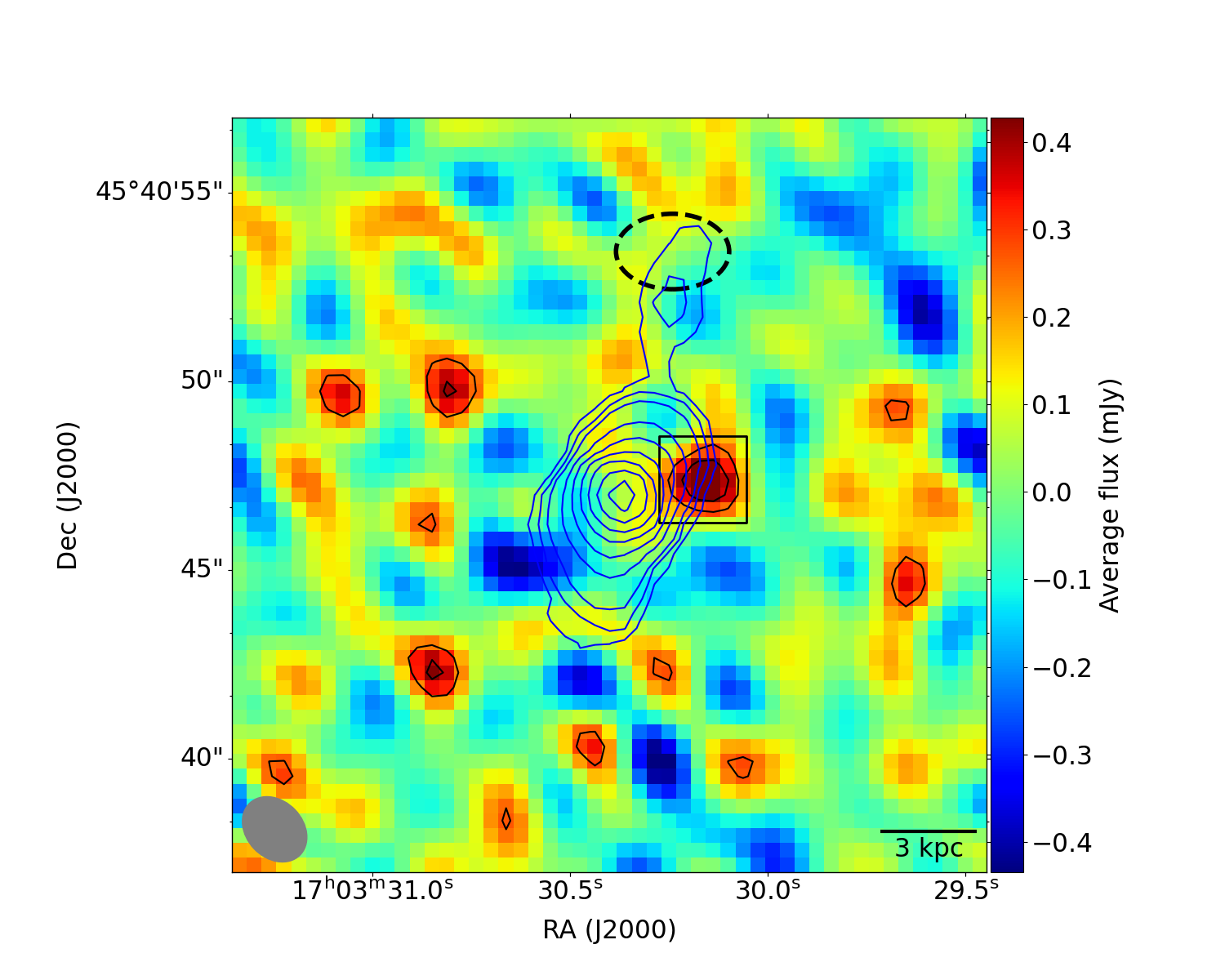}
  \hspace{3mm}
  \includegraphics[height=6cm,trim=15 25 50 50,clip=true]{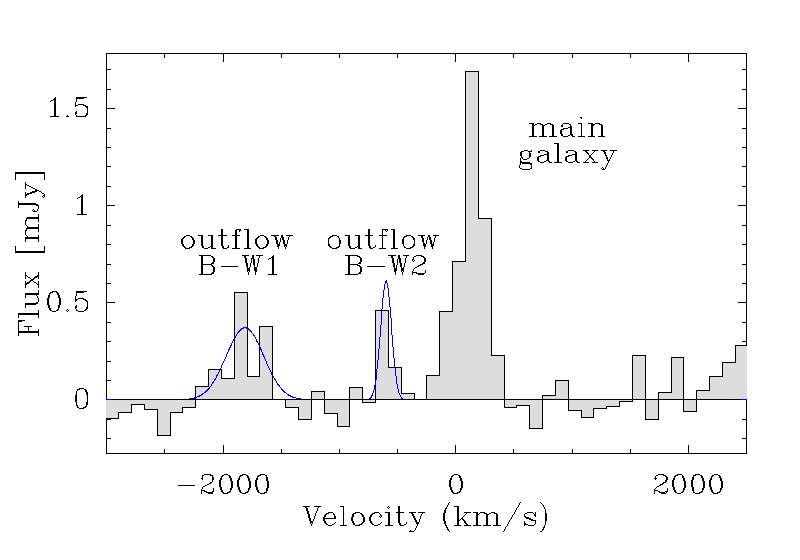}
  \caption{\emph{Left:} The color maps show the average  CO(1-0) line flux  obtained by averaging the visibilities in the velocity range $\sim$ -820 < v < -750\: km~s$^{-1}$ (\emph{top}) and $\sim$  -2000   < v <  -1300~km~s$^{-1}$ (\emph{bottom}). The black contours are 2,3,4 sigma. The overlapping blue contours show the systemic CO(1-0) emission from the host galaxy obtained by Salom\'e et al. (2021). \emph{Right:}  the integrated spectrum extracted from the region enclosed by the $2\sigma$ black contours in left panel (highlighted, for clarity, with a black open box).  These spectral plots are corrected by {\it z$_{CO}$} (i.e. same values as in Table~\ref{tab:noema}). The dashed, black line ellipse in the maps marks the position of the ``Northern Outflow" reported by Salom\'e et al. (2021).}
  \label{quentin_outflows_maps}
\end{figure*}

\begin{figure*}
  \centering
  \includegraphics[height=7cm,trim=0 0 85 0,clip=true]{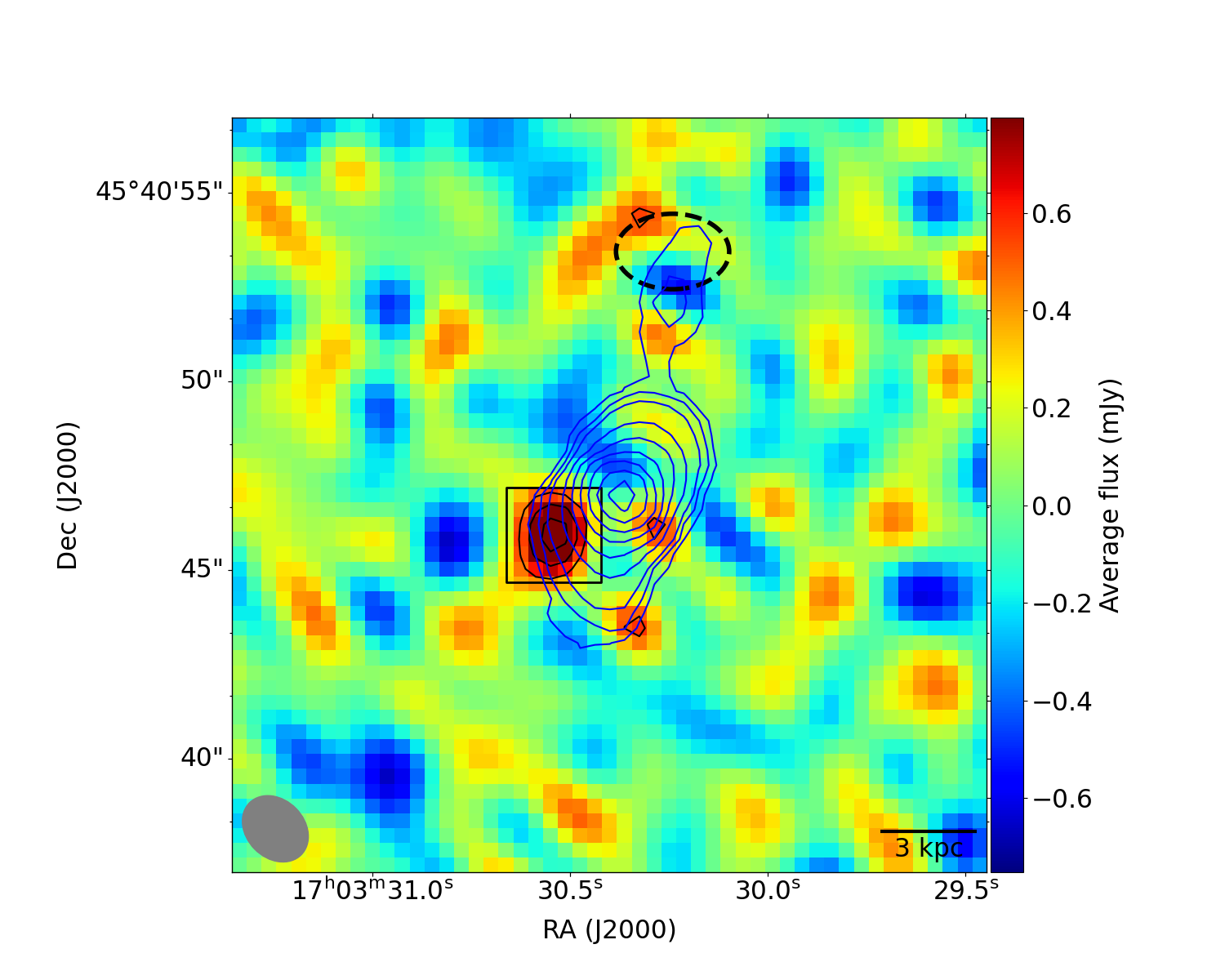}
  \hspace{3mm}
  \includegraphics[height=6cm,trim=15 25 50 50,clip=true]{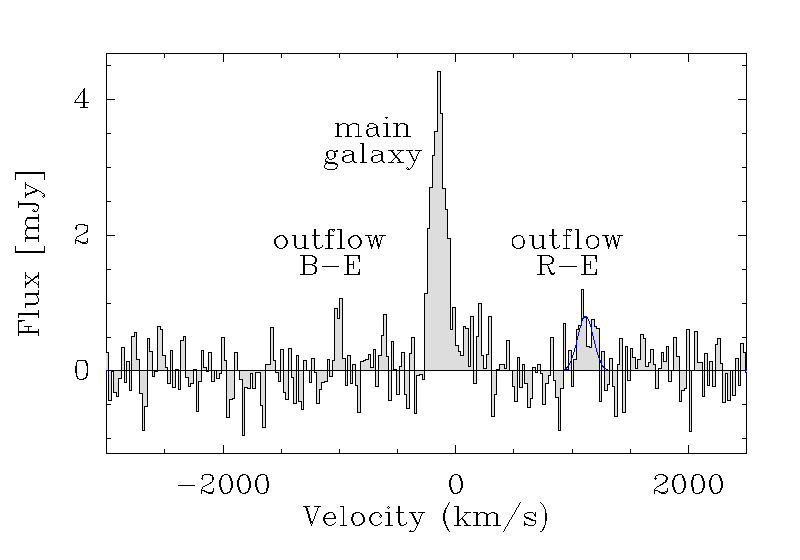} \\
  \vspace{3mm}
  \includegraphics[height=7cm,trim=0 0 85 0,clip=true]{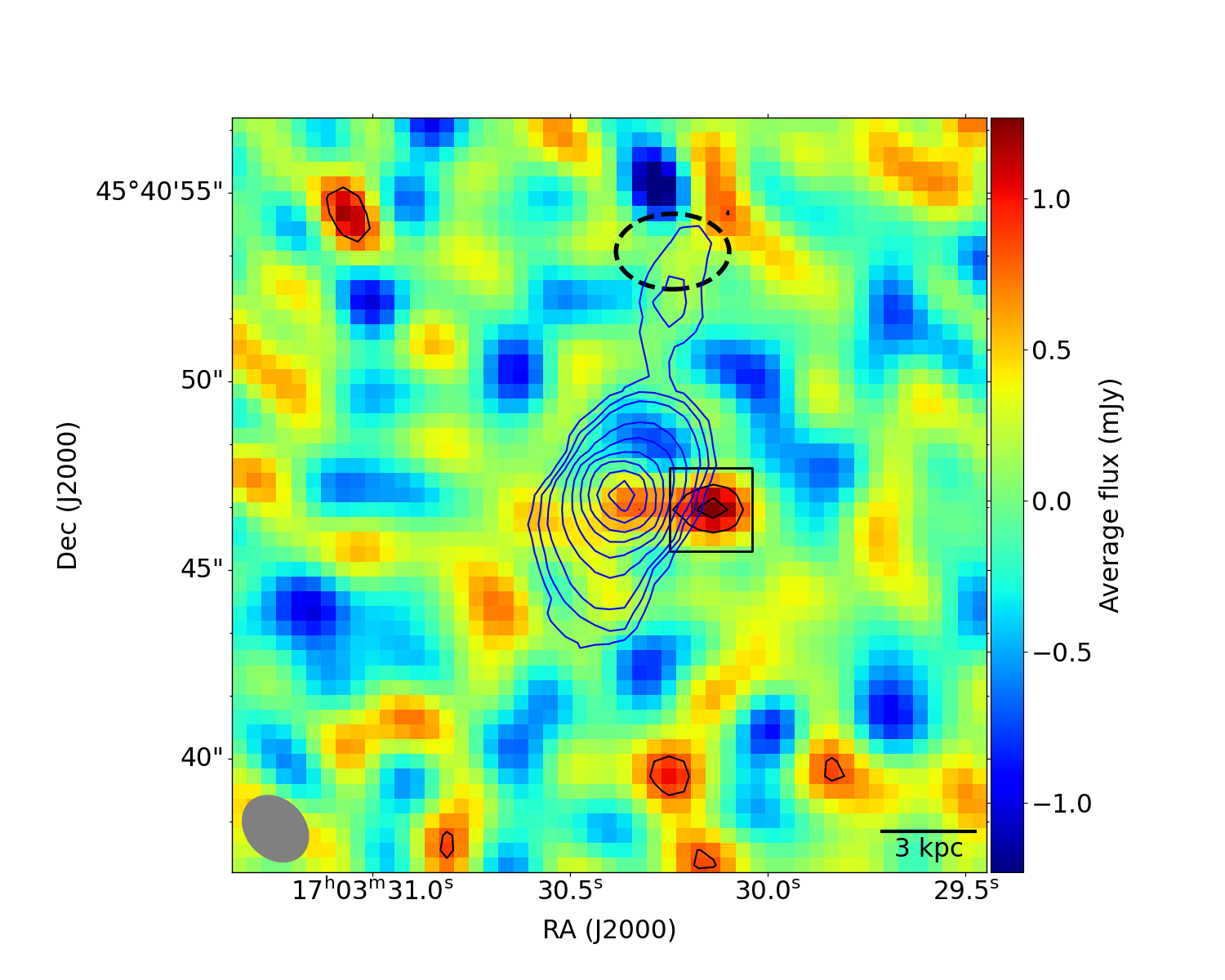}
  \hspace{3mm}
  \includegraphics[height=6cm,trim=15 25 50 50,clip=true]{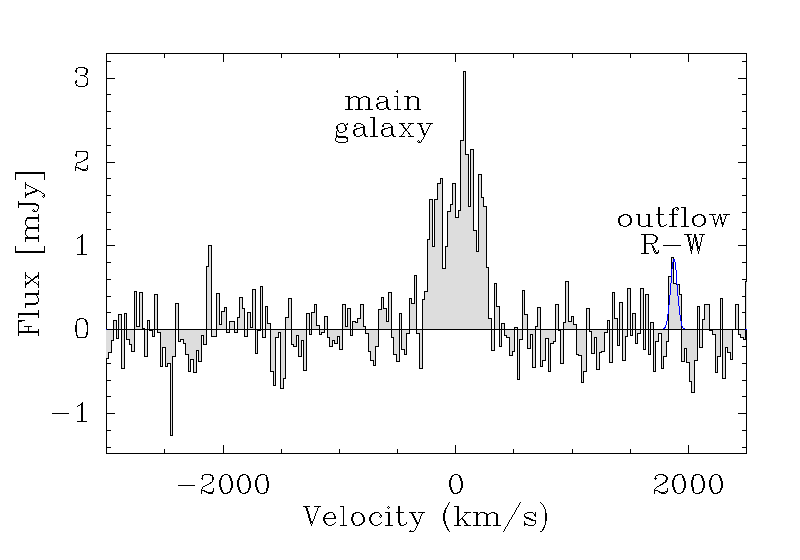}
  \caption{ \emph{Left:}  The color maps show the average  CO(1-0) line flux  obtained by averaging the visibilities in the velocity range  $\sim$ 1020 < v < 1255  km~s$^{-1}$ (\emph{top}) and $\sim$  1845 < v < 1925~km~s$^{-1}$ (\emph{bottom}). Everything else is the same as in Fig.~\ref{quentin_outflows_maps}.
  }
  \label{quentin_red_outflows}
\end{figure*}

Here we re-analyzed the NOEMA data cube produced by Salome et al. (2021). We started our exploration with a visual inspection of the molecular emission in the different velocity channels, taking as a reference the LMT broad CO line that peaks at $\sim$-660km~s$^{-1}$ and reaches values up to -1500 km~s$^{-1}$  (Longinotti et al. 2018).
Note that the velocities reported in this Section correspond to observed velocities, whereas the final values reported in Table~\ref{tab:noema} are corrected by {\it z$_{CO}$}. The maps obtained by averaging the visibilities in the negative velocity range $\sim$ -820 < {\it v} < -750\: km~s$^{-1}$ and $\sim$ -1300  < {\it v}  < -2000\: km~s$^{-1}$  show significant CO(1-0) emission in two distinct regions of the field. 
 The top panel of Fig.~\ref{quentin_outflows_maps} shows the integrated emission on the first velocity range. A spot of emission is seen slightly offset south-east from the nucleus  (``outflow B-E" hereafter).  The bottom panel of Fig.~\ref{quentin_outflows_maps} shows instead the integrated emission obtained in the higher velocity range: here, a second outflow (``outflow B-W1" hereafter) is detected west from the nucleus (bottom left map in Fig.~\ref{quentin_outflows_maps}).
The overlapping of the contours of the high velocity CO with the systemic molecular gas reported by Salom\'e et al. (2021) visually shows that the high velocity molecular gas is clearly separated from the galaxy rotation and that the outflowing gas is located  on a spatial scale of $\sim$ 3$^{\prime\prime}$, corresponding to a radial distance from the nucleus of about 3~kpc.

\begin{table*}
  \centering
  \caption{Parameters of the five outflows  components detected with NOEMA, along with the estimate of their combination (see \ref{subsec:combination}). The labels are the same as in Fig.~\ref{quentin_outflows_maps} and ~\ref{quentin_red_outflows}. The integrated fluxes are given by fitting a point source in the {\tt uv} table. We include a 10\% error to account for the calibration uncertainty. Peak velocities are corrected by {\it z$_{CO}$}.  
  The bottom row reports the properties of the ``Northern outflow" (see Section~\ref{subsec:outflow_N}) described in full details in Salom\'e et al. 2021. Molecular gas mass in the present paper is estimated assuming a conversion factor $\alpha_{CO}$=0.5~M$_{\odot}$~(K~km~s$^{-1}$~pc$^2)^{-1}$, with the exception of Outflow N, for which  Salom\'e et al. 2021 adopted a range  $\alpha_{CO}$=0.5--0.8 M$_{\odot}$ (K km s$^{-1}$ pc$^2)^{-1}$.}
  \begin{tabular}{lcccccccc}
    \hline \hline
    Outflow  &  $S_{CO}\Delta v$     &   $\Delta v$      &   $v_{peak}$         &   L$^{\prime}_{\rm{CO}}$ ($\times$10$^7$)  &    $M_{H_2}$  &  {\it R}  &   $\dot{M}_{H_2}$  &  $\frac{\dot{P}_{[CO]}}{\dot{P}_{rad}}$ \\
    component  & ($Jy\,km\,s^{-1}$)        & ($km\,s^{-1}$)  & ($km\,s^{-1}$)      &  [K km/s pc$^2$]                              &   ($10^7\: M_\odot$)  &   (kpc)      & ($M_\odot\,yr^{-1}$)  & - \\ \hline
                        
    {\bf B-W1} &   $0.39\pm 0.14$   &  $459\pm 80$   & $-1800\pm 35$   & 6.63$\pm$2.38                               &   $3.4\pm 1.2$                   &   3$\pm$0.3      &    $62\pm 20$     &  38$\pm$13   \\ %0.10+0.04
   {\bf  B-W2} &   $0.31\pm 0.10$   &  $137\pm 45$   &  $-610\pm 19$    &  5.27$\pm$1.70                             &   $2.7\pm 0.9$                  &    3$\pm$0.3     &  $16\pm 4$          &  3.3$\pm$0.9 \\ %0.07+0.03
        {\bf B-E}  &   $0.15\pm 0.05$   &   $67\pm 13$   &  $-1000\pm 10$  & 2.55$\pm$0.90                             &   $1.3\pm 0.4$               &   2.5$\pm$0.2   &     $16\pm 3$  & 5.3$\pm$1.1 \\ %0.03+0.02
       \hline 
    %    {\bf Total blue }   & - &    -&   $1136\pm500$  &  -&   7.4$\pm$2.5  &  3$\pm$0.3  &  85 [35 -- 163 ]   & 32 [7.4 -- 89]  \\ 
      %   {\bf  \scriptsize{(W1+W2+E)}} &  &   &  \\ 
 %   \hline\hline
     {\bf R-W} &     $0.10\pm 0.04$   &   $75\pm 15$   &  $1870\pm 20$  &    $1.7\pm 0.7$    &   $0.85\pm 0.35$   &  3.30$\pm$0.43 &    $15\pm 5$       & $10\pm 3$ \\
   {\bf  R-E}   &  $0.21\pm 0.07$   &  $150\pm 40$   &  $1100\pm 30$  &    $3.6\pm 1.2$    &    $1.8\pm 0.6$      &    2.09$\pm$0.43 &   29$^{+4}_{-5}$      &   12.0$^{+2}_{-3}$ \\
  % \hline 
%    {\bf Total red }  & - &    -&  1485$\pm$380 &  - &  2.65$\pm$0.95  &  2.7$\pm$0.4 &   44  [25 -- 66]  &  22 [9 -- 41]  \\ 
%{\bf  \scriptsize{(Red W+Red E)}}  & - &    -&   \\
  \hline\hline
        {\bf Total }   & - &    -&    $1280\pm480$  &  -&     10.0$\pm$3.5  &    2.8$\pm$0.3  &   139 [64 -- 234]   &   59 [17 -- 137]   \\ 
%    {\color{blue}      {\bf  \scriptsize{B\_(W1+W2+E)+R\_(W+E)}} }&  &   &  \\ 
            {\bf (All components) } &  &   &  \\ 

         \hline\hline 
    \bf{Outflow N} &  0.08$\pm$0.02        &   $45\pm 12$   &  $-790\pm 6$                                 &   $1.4\pm 0.4$                           & 0.9$\pm$0.3              &   7.8$\pm$0.3   &  2.7$\pm$0.9    & 0.8$\pm$0.3  \\ 
       \hline   \hline
  \end{tabular}
  \label{tab:noema}
 \end{table*}

   Both outflow components are unresolved by our NOEMA observation. We therefore fitted a point source within the averaged {\tt uv} tables to determine their positions. Outflows B-E and B-W1 are found at offsets of $[1.5,-1.4]$ and $[-2.5,0.5]$ arcsec, respectively. Following Reid et al. (1988), the position uncertainty $\Delta \theta$ on the position was estimated as
\begin{equation}
  \Delta \theta=0.5\frac{\lambda}{B}/SNR
\end{equation}
where $\lambda/B$ corresponds to the synthesised beam size  and SNR to the flux signal-to-noise ratio. The distance of the outflows of $2.1''\pm 0.2''$ and $2.5''\pm 0.3''$ respectively corresponds to 2.48$\pm$ 0.24\: kpc and 2.95$\pm$ 0.32\: kpc, which will be rounded to $2.5\pm 0.2$ and $3.0\pm 0.3$ kpc and assumed as the outflows distance  in the following.
The fit of a point source in the {\tt uv} plane also gives us information on the CO integrated flux (Table~\ref{tab:noema}). We respectively found 0.15$\pm$ 0.03\: Jy\,km~s$^{-1}$ for outflow B-E (corresponding to a 5.6$\sigma$ detection) and 0.39$\pm$ 0.10\: Jy\,km s$^{-1}$ for outflow B-W1 (4$\sigma$ detection).

   To determine the velocity of each outflow component, we extracted the integrated spectrum within the $2\sigma$ contours of the moment maps highlighted by the black box in Fig.~\ref{quentin_outflows_maps}. By fitting this emission with a Gaussian, we found that the emission in outflow B-E and B-W1 is respectively peaking at about -800 and -1600~km~s$^{-1}$. We note that the integrated fluxes given by fitting the spectral line are significantly smaller than the values obtained from the {\tt uv} table fitting that are quoted above. This is likely due to the low signal-to-noise ratio which complicates the deconvolution, therefore our results are derived by assuming  the integrated fluxes from the {\tt uv} fitting, as reported in Table~\ref{tab:noema}. 
   
In the integrated spectrum of outflow B-W1 (Fig.~\ref{quentin_outflows_maps} - bottom), a second contribution was detected with a peak velocity of $\sim$ -400\: km~s$^{-1}$. By fitting a point source to the average {\tt uv} table centred on this velocity, we found significant emission of outflowing gas  at an offset $[-2.2,0.5]$ and with an integrated flux of 0.31$\pm$ 0.07\: Jy\,km~s$^{-1}$ (corresponding to a 4.6$\sigma$ detection). This component is labelled B-W2. However, the low signal-to-noise ratio and the lack of spatial resolution do not enable us to separate B-W2 from the close-by outflow B-W1 in the map. 

 We then explored the data cube for the presence of significant CO(1-0) emission corresponding to receding gas motion. 
To this purpose, Fig~\ref{quentin_red_outflows} shows the maps obtained by averaging the visibilities in the positive velocity range $\sim$1020 < v < 1255  km~s$^{-1}$ and $\sim$  1845 < v < 1925~km~s$^{-1}$.  By applying the same procedure as above, significant CO(1-0) emission is detected in two areas located to the eastern and western side of the active nucleus, roughly at the same positions of the previously described ``blue components".

 Fitting a point source in the {\tt uv}  plane yields a further detection of two outflow components:  outflow R-E, with a significance of $4.6\sigma$ and a position offset of [1.47,-0.98]~arcsec that is  located at a distance of R=1.8$^{\prime\prime}$$\pm$ 0.3 (corresponding to $\sim$ 2.09$\pm$ 0.43~kpc) to the south-east side of the nucleus;   outflow R-W on the west side of the nucleus, which is detected at $3.7\sigma$  with an offset of [-2.77,-0.37]~arcsec (distance of  R=2.8$^{\prime\prime}$$\pm$0.3$\sim$ 3.30$\pm$0.43~kpc).
  The corresponding integrated  spectra  of these two ``red components" are plotted in the right panels of Fig~\ref{quentin_red_outflows}. As for the three blue components, outflow velocities were measured by fitting a Gaussian profile to these spectra.

Table~\ref{tab:noema} reports all parameters of the CO emission derived from {\tt uv} plane and spectral fitting, whereas the outflow physical properties are derived in the following section.
 Whenever in the following  ``the five outflow components" are cited, we refer to the outflows listed in the first five rows of  Table~\ref{tab:noema}, i.e. B-W1, B-W2, B-E, R-W and R-E. Remarkably, the spectral analysis of the CO in motion described herein shows that the velocity of these outflows is significantly higher than the value  (200-250~km~s$^{-1}$) reported by  Salom\'e et al. 2021 in their modeling of the rotating molecular disc of the host galaxy. This unequivocally  demonstrates   that the high-velocity CO gas, detected respectively with  $v_{out}$$\sim$1800~km~s$^{-1}$  and  $\sim$1000~km~s$^{-1}$ on the west and east side of the active nucleus,  is not consistent with the rotation pattern of the systemic molecular disc. 

Finally, we note that none of the outflow velocities reported for the NOEMA components  is corrected for projection effects:
   since this quantity, which is measured by the shift in the peak of the CO emission line, corresponds only to the velocity component of the gas moving along our line of sight, it is not possible to reconstruct what could be the de-projected  motion of the gas in other directions. 
 The same argument applies to the distance of the outflowing gas from the nucleus, which  is measured in the map as a projected radius: all NOEMA maps in Fig.~\ref{quentin_outflows_maps} and \ref{quentin_red_outflows} show that the outflowing  gas is located  within a volume at a distance $\ge$3~kpc  from the AGN nucleus, but the real position of each component relative to the others cannot be determined. For this reason, the measured velocities and distances shall be considered as lower limits.

\section{Results}
\label{sec:results}
 \subsection{Outflow mass output and energetics}

 \begin{figure}
 \includegraphics[width=1.\columnwidth]{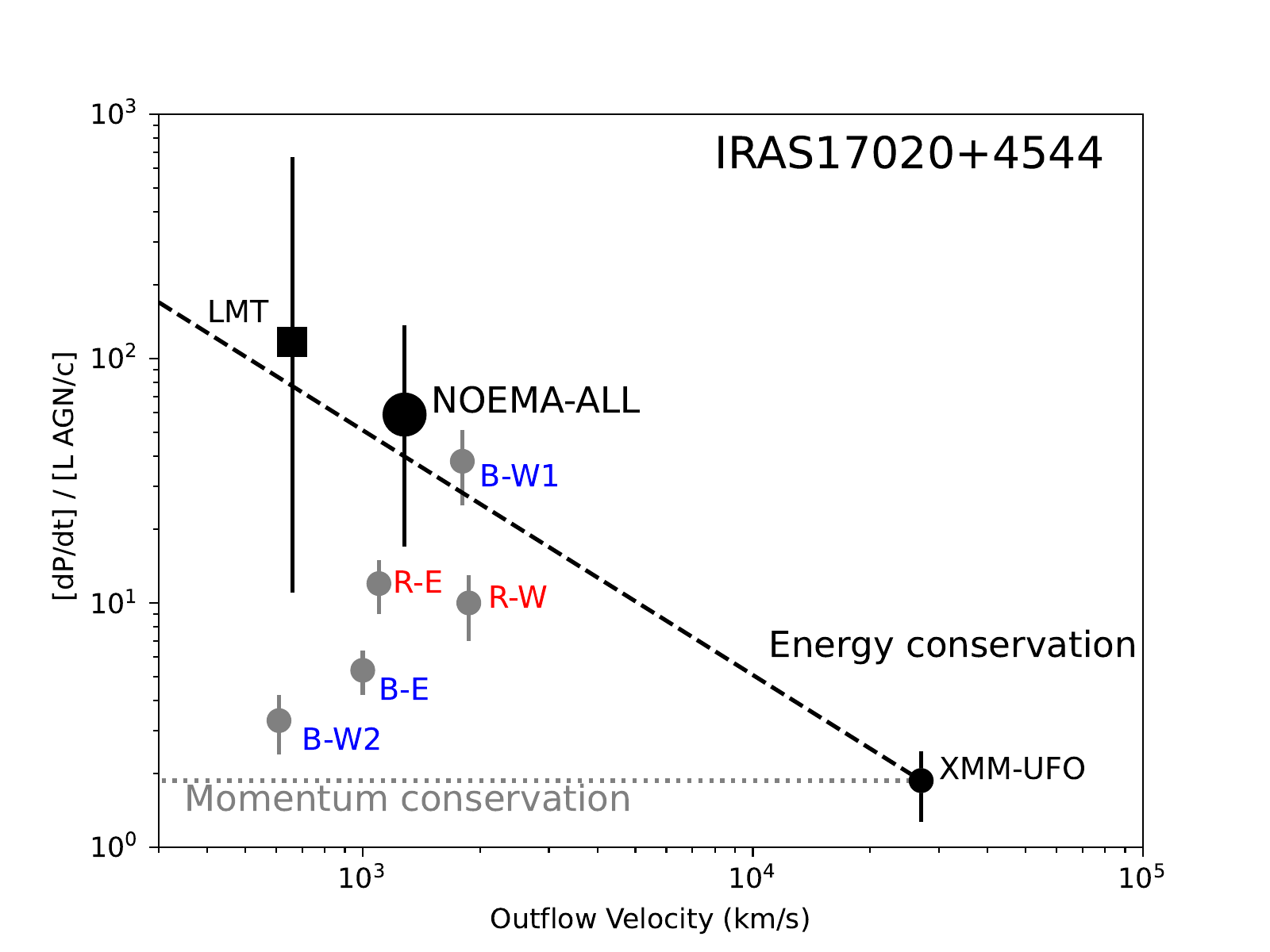}
 \caption{This plot  shows the relative force of the molecular and X-ray phases of the wind (expressed as the ratio of their momentum rates divided by the  AGN radiation momentum rate)
  plotted against their respective outflow velocities for IRAS17: gray circles mark the five NOEMA wind components reported in Table~\ref{tab:noema}. Red and blue labels for these points refer respectively to the receding and approaching velocities of the CO gas. The black circles mark the X-ray UFO from XMM-Newton (Longinotti et al. 2015) and the total molecular outflow obtained by combining the five NOEMA components.  The black square marks the former detection of the molecular outflow observed with the LMT  (Longinotti et al. 2018).
 The black dashed line marks the prediction for the  energy conserving outflow in the B-W1 component i.e.  
$\dot{P}_{[CO]}$/$\dot{P}_{[X]}$=v$_{out\_X}$/v$_{out\_CO}$. The grey dotted line marks the prediction for momentum-conserving outflows. }
\label{fig:ppunto_noema}
 \end{figure}

 The CO luminosity L$^{\prime}_{CO}$ estimated from the integrated intensity of each outflow component is reported in Table~\ref{tab:noema}. 
In the absence of more detailed knowledge of the molecular gas in this source, the mass corresponding to this CO luminosity  was estimated assuming  the lowest  CO-to-H2 conversion factor 
$\alpha_{CO}$=0.5~M$_{\odot}$~(K~km~s$^{-1}$~pc$^2)^{-1}$ reported for the  prototypical starburst  galaxy M82 (Wei{\ss}  et al. 2001). This factor, which is  almost 10 times lower than usually assumed for normal star forming galaxies (Solomon et al. 1987),  is the same adopted by Longinotti et al. 2018 in their LMT study of the CO properties of IRAS17, therefore this choice allows a direct comparison with previous results. However, an alternative choice is discussed in Section~\ref{subsec:comparison}.

The amount of mass that the outflow is expelling is estimated for each component under the assumption of spherical symmetry, therefore using the relation: 

\begin{equation}
\label{eq2}
\dot{M}_{[CO]}=3\frac{M_{CO} \times v_{out} }{R}
\end{equation}

As reported in Cicone et al. (2014) and extensively discussed in Harrison et al. (2018), the assumption of this relation implies that the outflowing gas is uniformly distributed within a region of spherical or multi-conical geometry, corresponding to a continuous flow of gas as opposed to an individual instantaneous event.

In this particular case, the higher spatial resolution of the NOEMA interferometer allows us to locate the molecular outflows on the spatial scale of 2.5-3~kpc (Table~\ref{tab:noema}), which is assumed as the radius $R$ in the above expression. 
Compared to the previous LMT detection, the outflow velocities are also measured with much higher precision in these data (see Table~\ref{tab:noema}), and we remark that all NOEMA velocity measurements herein reported are corrected for the shift of  225~km~s$^{-1}$  due to the  improved estimate of the source redshift based on the spectrum of the systemic CO gas   (Salom\'e et al. 2021).   In the following estimates, it is also important to keep in mind that since  the projection angle is not known, both measurements of the outflow distance and velocity are inevitably affected by this uncertainty. 

 Initially, in order to provide a detailed estimate of the mass and energy output released by the NOEMA outflow,  the five components will be treated separately.   
The resulting mass outflow rates estimated from Eq.\ref{eq2}  are respectively $\dot{M}_{[CO]}=16\pm 3$; $62\pm 20$;  16$\pm$4 M$_\odot$ yr$^{-1}$ for the ``blue outflows" B-E, B-W1, B-W2, whereas for the ``red outflows" R-W and R-E we respectively found $\dot{M}_{[CO]}=15\pm 5$; 29$^{+4}_{-5}$~M$_\odot$ yr$^{-1}$ (Table~\ref{tab:noema}).   

 The corresponding momentum rates of the wind can be estimated by $\dot{P}_{[CO]}$=$\dot{M}_{[CO]}$$\times$~{\it v}$_{[CO]}$. This yields 
 $\dot{P}_{[CO]}$=(6.4$\pm$2.2)$\times$10$^{35}$~cm~g~s$^{-2}$ for component B-W1 whereas lower values are found for the other four, being (0.91$\pm$0.18) (B-E), (0.56$\pm$0.15) (B-W2), (1.6$\pm$0.6) (R-W) and (2$^{+0.2}_{-0.5}$)  (R-E)  $\times$10$^{35}$~cm~g~s$^{-2}$ . 
 
 When the outflow momentum rate is compared with the AGN radiative momentum rate $\dot{P}_{{rad}}$~=~$\frac{L_{bol}}{c}$=1.7$\times$10$^{34}$~ cm~g~s$^{-2}$ (see Longinotti et al. 2018),  as summarized in Table~\ref{tab:noema}, their ratio  is found in the following proportion: $\frac{\dot{P}_{[CO]}}{\dot{P}_{rad}}$=38$\pm$13 (B-W1); 5.3$\pm$1.1 (B-E); 3.3$\pm$0.9 (B-W2); 10$\pm$3 (R-W) and 12$^{+2}_{-3}$ (R-E).  
 
We note that the ratio of the two forces (molecular outflow and radiative output) formerly estimated in the LMT data was a factor of $>$3 higher ($\frac{\dot{P}_{[CO-LMT]}}{\dot{P}_{rad}}$$\sim$117) compared to the value found for  the most powerful  NOEMA component B-W1.

Fig.~\ref{fig:ppunto_noema} shows the refined constraints provided by spatially resolving the five components of the outflow in the NOEMA data. By representing the momentum rate of each outflow normalized to the one of the AGN radiation as a function of its corresponding outflow velocity (see Fig.5 in Faucher-Giguere \& Quataert 2012), this plot describes the connection of the nuclear X-ray wind and the large scale molecular outflow. Upon energy conservation, the ratio between the momentum rate of the molecular and X-ray phases of the wind $\dot{P}_{[CO]}$/$\dot{P}_{[X]}$ is proportional to the ratio of their respective outflow velocities v$_{out\_[X]}$/v$_{out\_[CO]}$. 
In Fig.~\ref{fig:ppunto_noema} we also report the former estimate of the molecular outflow momentum rate  based on the LMT measurements and  the ratio of the X-ray wind momentum rate versus the radiative output of the source  $\frac{\dot{P}_{[UFO]}}{\dot{P}_{rad}}$=1.87$\pm$0.62. This number is the same reported by Longinotti et al. 2018 and we recall that it comes from the XMM-Newton measurement of the Ultra  Fast Outflow in IRAS17 (Longinotti et al. 2015). 

Out of the five outflow components  detected by NOEMA, outflow B-W1 is clearly  located  in the energy-conservation regime region of the diagram, showing a momentum boost  $\dot{P}_{[CO]}$/$\dot{P}_{[X]}$ of $\sim$20 with respect to the initial momentum retained within the nuclear X-ray wind.  On the contrary to B-W1,  the other components  are tracing an outflow regime where the momentum of the outflowing molecular gas undergoes a much  lower  boosting process. As discussed in Section~\ref{subsec:comparison}, this dichotomy is known in other AGNs although it is not frequently observed in outflow components of the same source.
We note that the error bars that constrain these  new measurements are significantly reduced compared to the LMT data point, mostly because the location of the outflow in NOEMA is extremely well constrained to within 3~kpc from the nucleus, as opposed to the wider range  (R$\sim$0.6-6~kpc) that was assumed to calculate the properties of the LMT outflow. The same applies to the NOEMA velocity measurements, which are determined with higher precision.  More details on the comparison of the LMT and NOEMA measurements are discussed in Section~\ref{subsec:compareLMT}.

\subsubsection{Combined properties of the AGN outflow}
\label{subsec:combination}
We followed Bischetti et al. (2019) to estimate the combined contribution of the outflow: the combined outflow velocity  was estimated as the mean of the five individual velocities estimated from the spectral fitting and  reported in Table~\ref{tab:noema}, whereas its errors were calculated as the standard deviation of the five velocities.  Although the total mass of the outflow reported in the sixth row of Table~\ref{tab:noema} is simply the sum of the five individual components, we decided to estimate the combined mass outflow  and momentum rates by applying the same methodology followed in the previous section, as opposed to summing the individual mass output listed in Table~\ref{tab:noema}.  Therefore these quantities are calculated according to Eq.~\ref{eq2} and assuming M$_{CO}^{comb}$=(10$\pm$3.5)$\times$10$^7$M$_\odot$ and v$_{out}^{comb}$=1280$\pm$480~km~s$^{-1}$.  The distance  of the ``combined outflow" was set to the averaged radius R=2.8$\pm$0.3~kpc.

The allowed ranges of the combined mass outflow rate and momentum rate are reported in the sixth row of Table~\ref{tab:noema}. 
 The boost in the momentum rate of the combined outflow compared to the one of the X-ray wind is   $\dot{P}^{comb}_{[CO]}$/$\dot{P}_{[X]}$=31.
The implications of these findings will be further discussed in Section~\ref{sec:discussion}.

\subsection{Comparison to LMT data}
\label{subsec:compareLMT}
 When the combined NOEMA outflow is considered (Table~\ref{tab:noema}), the total molecular gas mass in the five components  detected by NOEMA 
($\sim$1$\times$10$^8$~M$_\odot$) is a factor of $\sim$0.65 lower than the CO mass reported in the molecular outflow observed by the LMT, which in Longinotti et al. 2018  was estimated to 1.54$\pm$0.49$\times$10$^8$~M$_\odot$. However, it shall be noted that only the blue side of the outflow was measured in the LMT spectrum, therefore the combined mass of the NOEMA blue components (B-W1+B-W2+B-E), which corresponds to (7.4$\pm$2.5$)\times$10$^7$~M$_\odot$, should be directly compared to the LMT result.  Also, since NOEMA observations are sensitive to a spatial scale of 45$^{\prime\prime}$, the possibility of a more diffuse molecular outflow is unlikely, but we cannot exclude that extended emission is present and below the current sensitivity limit.

 The discrepancy in the mass of the outflowing gas can be understood by considering that the  NOEMA ``circumnuclear" outflows plus the Northern outflow component  (see Section \ref{subsec:outflow_N}) were blended in the larger LMT beam (see Fig.~3 in Salom\'e et al. 2021 to visualize the difference of the beam). 
It shall also be considered that the wide LMT error bars in Fig.~\ref{fig:ppunto_noema} include the uncertainty\footnote{In Longinotti et al. 2018, the measurement in this diagram is estimated from a CO mass obtained with a conversion factor spanning the range $\alpha_{CO}$=0.5--0.8 M$_{\odot}$ (K km s$^{-1}$ pc$^2)^{-1}$ } on the molecular gas mass, whereas the  CO mass embedded in the NOEMA points, is derived with $\alpha_{CO}$=0.5 M$_{\odot}$ (K km s$^{-1}$ pc$^2)^{-1}$ (see Table~\ref{tab:noema}).  Overall, the new constraint provided by the NOEMA data shows that the molecular outflow properties measured by the two millimeter observatories are consistent within the errors.

%\begin{figure}
%  \includegraphics[width=1.1\columnwidth]{fig10.png}
%   \includegraphics[width=1.1\columnwidth]{fig11.png}
% \caption{\emph{Top:} Composite image of IRAS17, the spatial scale is 1.18kpc/$^{\prime\prime}$: blue and red contours trace the approaching and receding molecular outflowing gas reported in Section \ref{mol_out}, the label ``N" marks the position of the Northern Outflow; grey contours trace the systemic CO gas in the galaxy reported in companion paper (Salom\'e et al. 2021). The grey scale underlying image is the {\it Chandra} X-ray emission as observed by the HRC camera (see Section~\ref{subsec:xray}), green contours trace the radio emission observed by  e-MERLIN at 1.5~GHz (see Section~\ref{subsec:radio}). The white cross at the center of the image marks the {\it Gaia} position of the optical AGN. The upper ``node" detected in the systemic CO  emission close to the  Northern Outflow position, corresponds to the position of the companion galaxy. \emph{Bottom:} Zoom of the area enclosed by the dashed line in the upper panel. The position of the molecular outflows were labelled with letters with the same colour code as above. Note that due to the overlapping of  B-W1 and B-W2, a single blue label ``W" is adopted to mark these two components. }
% \label{fig:Chandra}
%\end{figure}

  \subsection{The host galaxy and the  Northern Outflow}
  \label{subsec:outflow_N}

\begin{figure}
  \includegraphics[width=1.1\columnwidth]{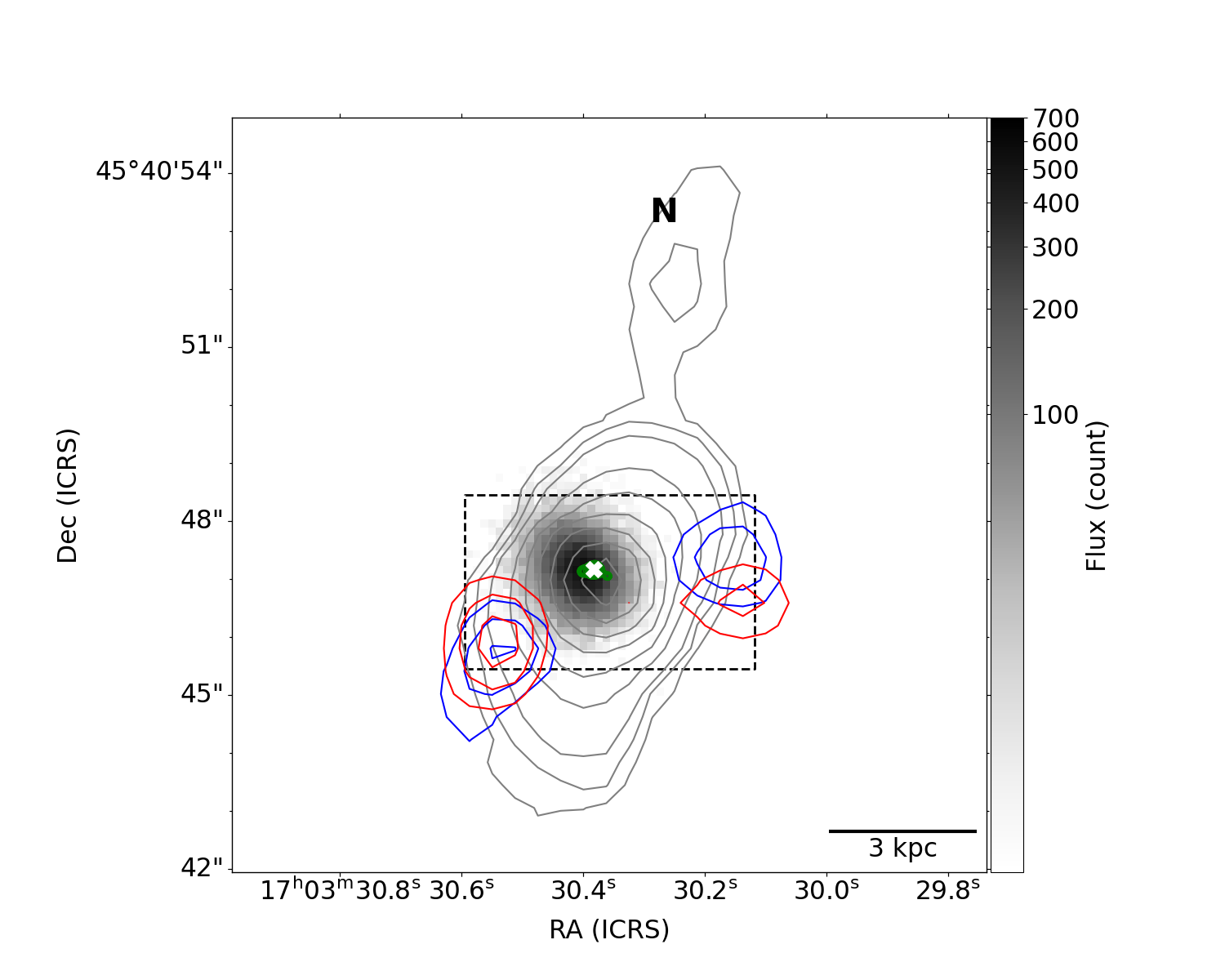}
   \includegraphics[width=1.1\columnwidth]{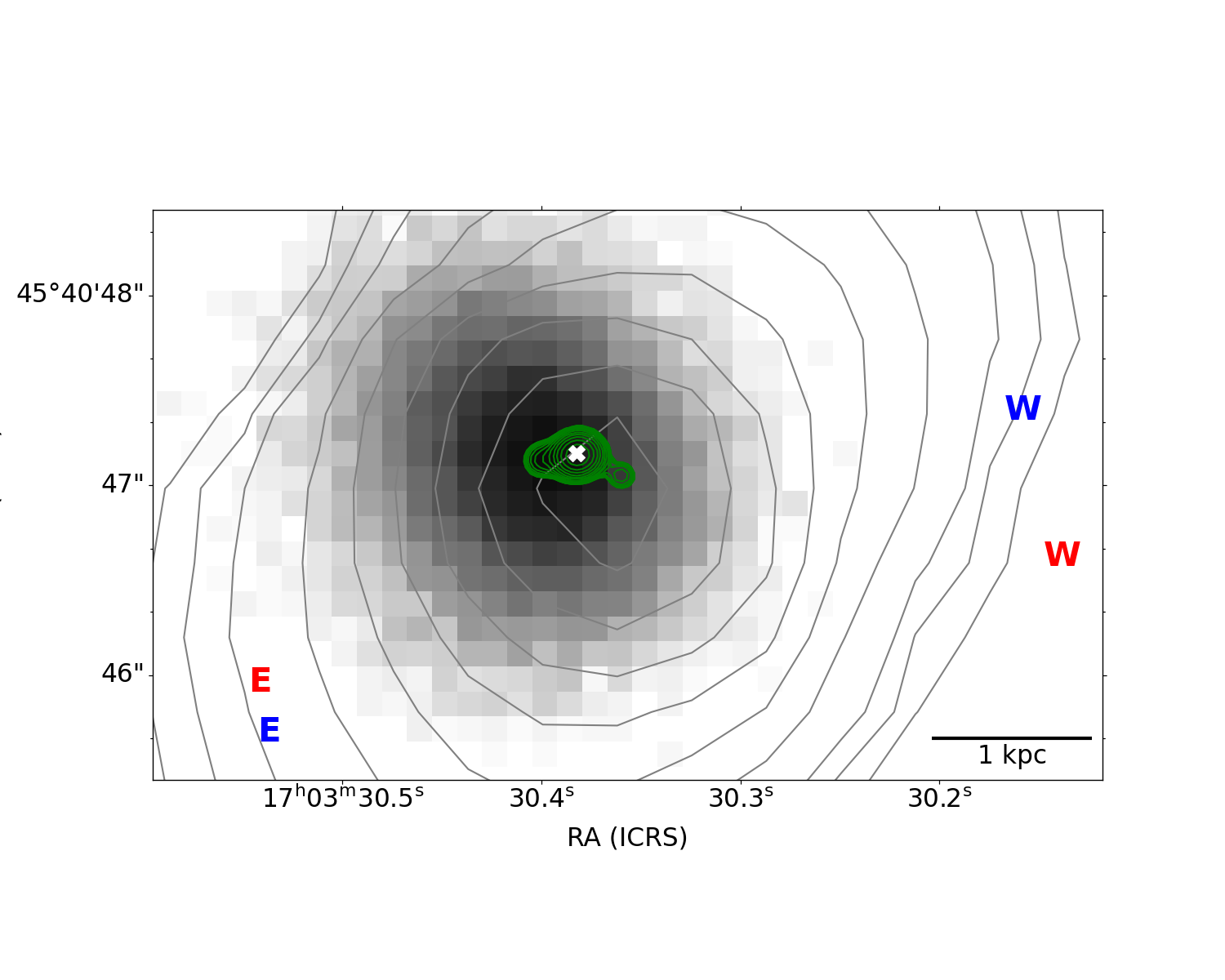}
 \caption{\emph{Top:} Composite image of IRAS17, the spatial scale is 1.18kpc/$^{\prime\prime}$: blue and red contours trace the approaching and receding molecular outflowing gas reported in Section \ref{mol_out}, the label ``N" marks the position of the Northern Outflow; grey contours trace the systemic CO gas in the galaxy reported in companion paper (Salom\'e et al. 2021). The grey scale underlying image is the {\it Chandra} X-ray emission as observed by the HRC camera (see Section~\ref{subsec:xray}), green contours trace the radio emission observed by  e-MERLIN at 1.5~GHz (see Section~\ref{subsec:radio}). The white cross at the center of the image marks the {\it Gaia} position of the optical AGN. The upper ``node" detected in the systemic CO  emission close to the  Northern Outflow position, corresponds to the position of the companion galaxy. \emph{Bottom:} Zoom of the area enclosed by the dashed line in the upper panel. The position of the molecular outflows were labelled with letters with the same colour code as above. Note that due to the overlapping of  B-W1 and B-W2, a single blue label ``W" is adopted to mark these two components. }
 \label{fig:Chandra}
\end{figure}

  Fig.\ref{fig:Chandra} shows the composite image obtained by combining  the spatial information of the five molecular outflow components with the multi-band view of IRAS17 accumulated over the recent years. The gray contours in this image mark the systemic CO gas associated to the host galaxy rotation studied by Salom\`e et al. 2021. Red and blue colours respectively mark the receding and approaching  sides of the molecular outflows (both contours and letters marks) following the same nomenclature as in Fig.~\ref{quentin_outflows_maps}, Fig.~\ref{quentin_red_outflows} and Table~\ref{tab:noema}; the emission in gray scale distributed around the centre of the image traces the X-ray source as seen by {\it Chandra}  (see Section \ref{subsec:xray}); green contours mark the position and extension of the radio source in the 1.5~GHz observed by e-MERLIN (see Section \ref{subsec:radio}); finally, the white cross, marks the position of the optical AGN observed by {\it Gaia}. 
  
   The upper ``node" in the CO gas distribution that  in the top panel of Fig.\ref{fig:Chandra} is clearly located out of the galaxy disc, marks the companion galaxy discovered by  Salom\'e et al. (2021). 
These authors also reported the detection of an outer outflow component, which is marked by label ``N" in the top panel of Fig.~\ref{fig:Chandra}, located in projection of the companion galaxy discovered in the NOEMA data (see Fig. 8 and Table 5 in their paper). The properties of this outer wind, which we refer to as ``Northern Outflow", are listed in the bottom row of Table~\ref{tab:noema}.
Although the velocity of this component (-790$\pm 6$ km~s$^{-1}$) is similar to some of  the values measured for the nuclear outflow components, the lower molecular gas mass  of (0.9$\pm$0.3)$\times$10$^7$~M$_\odot$  and   the higher distance from the nucleus, R=7.8$\pm$0.3~kpc, yield a moderate mass output  $\dot{M}_{[CO]}$= 2.7$\pm$0.9 M$_\odot$ yr$^{-1}$ compared to the circumnuclear outflow. 
The peripheral position of this outflow makes it less likely for  this outflowing gas to bear relation with the current AGN activity and to the circumnuclear molecular outflow, rather, it is probably connected with the merging activity, therefore it is not further discussed nor is it included in the following figures and tables. 

Additional information on the Northern Outflow and the companion galaxy will be provided by the ongoing analysis (Robleto-Or\'us et al. in prep.) of the spatially resolved optical spectra obtained by our group with the Integral Field Unit instrument MEGARA on the GTC Telescope (Gran Telescopio de Canarias).

\section{Discussion}
\label{sec:discussion}

The latest observational campaign carried out with the NOEMA interferometer and hereby reported, confirms that the NLSy1 IRAS17 is  characterized by a truly multi-phase outflow.
\subsection{What do we know about IRAS17}
Before going through the discussion of the various aspects of this campaign and with the aim to provide the reader with a ready-to-use overview on the source,  we  summarize in the following the main findings on this peculiar AGN provided by past, present and ongoing  publications. 
\label{subsec:list_properties}
\begin{itemize}
\item A powerful molecular outflow resolved in  five components extending on a scale of $\sim$3~kpc from the active nucleus is revealed by observation of the CO(1-0) emission with the NOEMA interferometer (this paper).  The spatial distribution of the outflowing molecular gas (Fig.~\ref{fig:Chandra}) is consistent with  the presence of a biconical structure located in projection on each side of the nucleus, and respectively traced by the receding and approaching gas motion.   The ``Western outflow" is traced by components B-W1 and R-W, both reaching a line of sight velocity of around 1800-1900~km~s$^{-1}$, whereas  the ``Eastern outflow",  traced  by B-E and R-E, appears to be outflowing at slightly lower velocities ($\sim$1000~km~s$^{-1}$),  although these velocities are to be considered as lower limits due to projection effects.
The  mass  expelled by the combined outflow, including also the B-W2 component, is estimated to be  $\dot{M}_{[CO]}$$\sim$139~$M_{\odot}$yr$^{-1}$ and its momentum rate exceeds the one of the AGN luminosity by a factor of almost 60.  The boost in the momentum rate of the combined outflow compared to the X-ray nuclear wind is then of a factor of $\sim$30. This result leads to a solid confirmation of  the energy-conserving nature of this large-scale wind. 
\item The systemic CO gas in the NOEMA data (Fig.~\ref{fig:Chandra}, and Salom\'e et al. 2021) traces the  rotation of the molecular gas that is distributed in a well-organized pattern typical of a galaxy disc/ring. Interestingly, Salom\'e et al. have also uncovered the existence of a companion galaxy of $\sim$4.5~kpc of diameter and molecular mass of about 10$^8$$M_{\odot}$ located to the north of the main active nucleus and within the northern extension of the systemic CO. This discovery shifts the classification of IRAS17 from a classical spiral Seyfert galaxy to a minor merger system. The lack of disturbed dynamics in the CO emission and in optical images led Salom\'e et al. to postulate that the two galaxies are still at an early phase of a merger.  
\item Prior X-ray spectroscopy provided by {\it XMM-Newton} and {\it Chandra} revealed a wealth of information on the nuclear wind:  i) a powerful X-ray stratified ultra-fast outflow (Longinotti et al. 2015) where the  momentum rate of the most massive component exceeds the momentum output of the AGN radiation ($\frac{\dot{P}_{[X]}}{\dot{P}_{rad}}$=1.87$\pm$0.62);  ii) a multi-layer warm absorber-type wind (Sanfrutos et al. 2018) with components that over a 10 yr time scale appear to be in outflow (velocities within 2-4$\times$10$^3$~km~s$^{-1}$), in inflow (velocities within 1.5-3$\times$10$^3$~km~s$^{-1}$) and stationary ($\sim$400~km~s$^{-1}$). The complex ensamble of the warm absorber and the ultra fast outflow, is confirmed in {\it Chandra} LETG-spectroscopy (Longinotti 2020, and paper in prep.). 
\item  
Preliminary results of ancillary radio and X-ray imaging included in Fig.~\ref{fig:Chandra} tentatively show that: 1) diffuse X-ray emission cannot be excluded in the nuclear region of the galaxy (see Section~\ref{subsec:xray}), with an elongated shape towards the NE side of the nucleus; 2) the presence of a radio source on the arcsec scale composed by a  dominant central core with two secondary components aligned on the east  and south-west side of the core (see Section \ref{subsec:radio});   3) the e-MERLIN emission accounts for the majority, but not the entirety, of the total radio emission detected, e.g., in the NVSS and FIRST surveys at 1.4 GHz (see Giroletti et al. 2017). This leaves room for some additional diffuse components on larger scales or to significant variability from the nuclear region. 
\item Beside the radio loudness of this source and the well-established presence of synchrotron radiation in the compact core (Giroletti et al. 2017), radio observations of IRAS17 at various frequencies and spatial scales (see J{\"a}rvel{\"a} et al. 2022 for a summary of radio properties), have highlighted the presence of diffuse radio emission located within the central kpc, whose power cannot be entirely explained by star formation processes.
\item   A recent analysis of the {\it HST-COS} IRAS17 spectrum reported  the presence of the Ly$\alpha$ line in absorption in the Ultraviolet band  consistent with being the counterpart of the low ionization X-ray UFO (Mehdipour et al. 2022).  The presence of a UV fast outflow in the source is interpreted as the effect of the X-ray UFO entraining and shocking  the cold and/or warm  Inter Stellar Medium (ISM) of the host galaxy. 
\end{itemize}

These considerations naturally prompt some questions on the molecular outflow origin: 
\begin{enumerate}
\item Were the five NOEMA components expelled at the same time or are we witnessing different episodes of outflows? 
\item Does their relative orientation with respect to the AGN nucleus and to the radio emission tell us something on the mechanism of propagation, and therefore on their capability of conserving the initial energy released by the nuclear X-ray wind? 
\item Was the structure of the surrounding ISM  somehow responsible for shaping and channeling each outflow component in the status as we observe it now? 
\item Since the energy is efficiently transported from the vicinity of the black hole to outer regions, what is the impact of the outflow to the rest of the galaxy?
\end{enumerate}
We attempt to address these important questions and to provide a working scenario of the outflowing gas in IRAS17 in the next sections, where the molecular outflow is discussed along with the other multi-band aspects of this AGN.

\subsection{The NOEMA outflow}
\label{subsec:noema_outflows}

The novel results provided by the analysis of the NOEMA data have uncovered a fairly unexpected scenario in this Seyfert 1 galaxy.   First of all, the massive molecular outflow originally revealed in the single-dish data by LMT is now clearly resolved in multiple components (see Table~\ref{tab:noema} and Fig.~\ref{fig:Chandra}), allowing geometrical considerations to be made. 

The location and the properties of the outflow with its Eastern and Western sides reported in  Section~\ref{mol_out} and Fig.~\ref{fig:Chandra} suggest to envisage in IRAS17 a standard  biconical 
geometry for the outflowing molecular gas, similar to what was proposed for other sources, e.g.  Mrk 231 (Feruglio et al. 2015), NGC4945 (Bolatto et al. 2021). 
In these sources, high-resolution mapping of the molecular gas has allowed a very fine tracing of the hollow cones and of their position relative to the galaxy molecular disc, which is not possible to replicate with the current  NOEMA dataset due to the lower spatial resolution and signal-to-noise. 
 On the contrary to these sources where the molecular outflow includes residual rotation from the galaxy disc and is located close to the nucleus, 
 the outflow in IRAS17 with its velocities ranging at least from  $\sim$1000 to $\sim$1800~km~s$^{-1}$ is clearly characterized by a more extreme dynamical behaviour, also in comparison to other literature results (see Lutz et al. 2020). The observed high velocity  allows us to exclude residual contributions from the rotating disc, whose velocity field reaches a maximum of $\sim$200-250~km~s$^{-1}$ (Salom\'e et al. 2021), also in the case of the  lower velocity outflow component (B-W2, with v$_{out}$ of about -610~km~s$^{-1}$).

Moreover, the detection of redshifted  and blueshifted outflow components in each side of the cone at an extremely high velocity,  provides further clues to interpreting the outflow phenomenology at galaxy-scale. Such structure is indeed  highly reminiscent of the kinematic pattern of ionized and atomic gas observed in great details in many Seyfert galaxies and interacting systems (e.g. M{\"u}ller-S{\'a}nchez et al. 2011, Mingozzi et al. 2019, Perna et al. 2019).  
This peculiar velocity structure may well be the result of the  action of the inner X-ray Ultra-Fast-Outflow that is shocking and compressing the ISM (see Section \ref{sec:shocked_outflow}).

With regard to the effect that the outflow may have on the galaxy molecular disk,  we note that the mass measured by NOEMA in the most massive component (B-W1) of the CO outflow is  ($3.4\pm$1.2)$\times$10$^7$~$M_{\odot}$ (Table~\ref{tab:noema}). When compared to the total CO mass measured in the systemic gas reported by Salom\'e et al. 2021 (M$_{[CO]}$=$\sim$1.1$\times$10$^9$~$M_{\odot}$), this implies that this  component  ejects on its own almost a 3\% of the molecular gas available in the galaxy disc.  
   The combined mass contained in the outflow (see Section \ref{subsec:combination}) is estimated to $(10\pm$3.5)$\times$10$^7$~$M_{\odot}$, which corresponds to $\sim$10\% of the total mass of the systemic CO gas in the host galaxy disc.

\subsection{The multi-phase outflow and its host galaxy: clues from the X-ray and radio images} 
\label{subsubsec:spatial}

 \subsubsection{The elusive properties of the radio source in IRAS17}
 \label{sec:radio}
As to the radio properties, the  steep  spectrum  and  the  high-brightness  temperature of  $\sim$10$^{7-8}$~K  reported by Giroletti et al. (2017), point  unambiguously  to  a synchrotron origin of the radio emission on pc-scale. However, from that analysis, the non-thermal plasma responsible for the synchrotron radiation did not appear to be in bulk relativistic motion, based on the tight upper limits on the presence of significant advance motion in the parsec scale features, as well as based on the lack of any other typical signatures of Doppler beaming (see Giroletti et al. 2017).  It thus remains unclear whether the radio emitting plasma is accelerated as the consequence of the outflow shocking with the ISM or in processes more typical of the classical collimated jets of radio-loud AGNs (see Panessa et al. 2019 for a discussion on these mechanisms). The apparent absence of motion and of classical jets characteristics, along with the outflow  properties listed in Section~\ref{subsec:list_properties} and further elaborated  in Section \ref{sec:shocked_outflow}, led us to consider the intriguing hypothesis that the synchrotron emission could well be arising via  accelerated electrons in shock processes (e.g. Nims, Quataert \& Faucher-Gigu\`ere, 2015). 

The sub-arcsec radio elongated structure seen in the e-MERLIN image (see Section~\ref{subsec:radio}
 and Fig.~\ref{fig:emerlin}) extends approximately on a scale of 1~kpc  with an apparent  orientation along the East-West axis, similar to the one of the  CO outflow. Projection effects may be at work also in this case. Deriving a more precise  physical link between the radio and the molecular gas emission is not possible based on the existing data. To this purpose, deeper radio observations on a larger scale, which may  ``bridge" the current  gap between the spatial scales of the radio plasma and the molecular gas are needed.   The archival VLA data of IRAS17 reported by Berton et al. (2018), which are sensitive up to a 10~kpc scale, show  compact radio emission  (2-300~pc)  and only an apparent E-W elongation that is likely due to the convolving beam and the short VLA integration time (9 minutes).

Nonetheless, we could estimate that the VLBI images (Giroletti et al. 2017) only recover $\sim 60\%$ of the VLA unresolved flux density, giving a solid indication for  the presence of significant radio emission on more extended scales that could well encompass  the molecular outflow traced by NOEMA.  Similar conclusions were reached by J{\"a}rvel{\"a} et al. (2022), who, in their JVLA study, effectively propose that non-thermal mechanisms are responsible for the observed  radio emission.  If eventually revealed, such extended radio plasma has a profound impact on the surrounding ISM, as postulated by simulations of interaction of compact jets and ISM (e.g. review by Mukherjee et al. 2021 and references therein), and, ultimately,  it may as well have a key role in driving the outer molecular outflow, as revealed in other mildly radio-emitting  Seyfert sources (e.g. Husemann et al 2019, Morganti et al 2015).

The study of the radio properties in sources with galaxy-scale and nuclear outflows is indeed a promising field: recently,  parsec-scale jetted structures of highly accreting active galaxies  have been revealed in VLBI data (e.g. in the Seyfert 1 Mrk 335, Yao et al. 2021; in the changing-look AGN Mrk 590, Yang et al. 2021; in the powerful QSO PDS456, Jun et al. 2021). Although the link of these jets with the nuclear and large-scale activity is still elusive (as in the case of IRAS17),  the radio emission has great potential  for pinpointing the mechanism that drives the energy release into the ISM.

 \subsubsection{Molecular and X-ray gas distribution}
As detailed in \ref{subsec:xray} and shown in Longinotti et al. 2022, a tentative extension of the X-ray hot gas seen by {\it Chandra} seems to exist in orthogonal direction (North-East of the nucleus) with respect to the cone traced by the CO outflow.

Most shocked outflow models predict extended X-ray emission in the 1-10~keV band produced in the ambient medium  swept-up by the forward shock via Inverse Compton and free-free emission mechanisms (e.g. Nims, Quataert \& Faucher-Gigu\`ere, 2015). The X-ray gas emitted by the shocked wind is  predicted to be in excess compared to what is expected by  normal star formation  processes that also produce diffuse X-ray emission (Mineo et al. 2014). Unfortunately, as explained in Section~\ref{subsec:xray}, the putative spatially extended X-ray signature of the shocked wind in Fig.~\ref{fig:Chandra} cannot be disentangled from the AGN nuclear continuum, therefore solid conclusions cannot be drawn from the current {\it Chandra} image. Further speculations on the X-ray extended emission are discussed in Section~\ref{sec:bubble}.

 Nonetheless, the relative orientation of the galaxy disc traced by the rotation of the systemic CO gas in Fig.~\ref{fig:Chandra} is also telling important information on the possible geometry of the multi-phase outflow.  
Salom\'e et al. 2021 reports that this gas is distributed in a disc of radius of about 3.5~kpc.  The rotation curve in their paper was modeled with an inclination of 60-70$^{\circ}$, consistently with optical imaging that suggests that the galaxy is seen  almost edge-on (see Fig.2 and Fig.3 of Salom\'e et al. 2021). These figures along with the aforementioned analysis  also show that the molecular disc is contained within the bulge of the galaxy, and that it is not affected by perturbation.
Therefore, we can expect that the CO outflow and the X-ray putative diffuse emission are expanding normal to the galaxy disc. We elaborate on this idea in the next Section.

\subsection{Energetics of the X-ray and molecular winds: a shocked outflow in IRAS17}
\label{sec:shocked_outflow}
The observational facts outlined in Section~\ref{sec:discussion}  can be coherently framed if we postulate that the nuclear, ultra fast X-ray wind is the ``driver" of a shock expanding within the central kpc of the galaxy. The theoretical framework to this idea is provided by several analytical models: e.g. King (2010), Faucher-Gigu\`ere \& Quataert (2012), King et al. (2011). 
We will show  in the following that all ingredients described in these models are indeed present in the observational history of IRAS17.

From X-ray spectroscopy, we know that the ultra fast outflow launched with an outflow velocity of $\sim$30,000km~s$^{-1}$(Longinotti et al. 2015)  produces  various components of the wind organized to span a wide range of ionization and column density that may well arise in the shock originated by the impact of the fast wind with the ISM. This process could be understood by considering that the difference in the gas density in front and behind the shock discontinuity is likely to impart Rayleigh-Taylor instabilities as in Supernova Remnants (see Velazquez et al. 1998, Zubovas \& King 2014), that would result in ``sections" of gas with different conditions (velocity, ionization and column density).  When crossing our line of sight, these ``gas sections"  would be seen as wind components with a stratified structure in the X-rays, as observed in X-ray spectra of IRAS17 (Sanfrutos et al. 2018 and Longinotti 2020).

This phenomenology and its interpretation shall not be seen as  exclusive to IRAS17: in recent years, the observation of stratified UFOs has become more and more frequent in bright Seyfert Galaxies (e.g. Reeves et al. 2020, Krongold et al. 2021, Xu et al. 2021), suggesting the existence of a common and established mechanism  that in some AGNs provides a coupling of  the UFO with the surrounding ambient gas of the galaxy. 
In this framework, the detection in IRAS17 of the UV counterpart (Mediphour et al. 2022) of one of the low ionization X-ray UFO brings in additional evidence for entrainment of the shocked ISM by the ultra fast wind, as proposed by Serafinelli et al. (2019).

However, the most compelling ingredient for the shocked outflow interpretation is provided by the properties of the molecular outflow revealed by NOEMA, namely the energy conservation and the presence of high velocity red- and blueshifted components. 
While the energy-conserving nature of the outflow in IRAS17 was first postulated based on LMT data, the outflow location and geometry could not be directly determined from these single-dish observations, which lacked  spatial information (Longinotti et al. 2018).
The present  interferometric data have allowed us to resolve the different components and therefore to pinpoint the properties of the outflowing molecular gas in an unprecedented way.
 In terms of energetics, Fig.~\ref{fig:ppunto_noema} shows that while B-W1 is undoubtedly located in the energy-conservation regime ($\dot{P}_{[CO]}^{B-W1}$/$\dot{P}_{[X]}$ $\sim$20), the other four outflow components  are consistent with having received a much lower boost to their initial momentum. 
 Remarkably, the combined properties of the outflow (Section \ref{subsec:combination}) are still fully consistent with energy conservation.  
 The global molecular outflow seen by NOEMA provides therefore a solid confirmation that the initial energy of the wind is efficiently transferred outward during its propagation through the galaxy, as expected by the action of an expanding shocked wind (e.g. Faucher-Gigu\`ere \& Quataert 2012).
Assuming an average distance of $\sim$3~kpc, the combined mass M$_{CO}^{comb}$=(10$\pm$3.5)$\times$10$^7$M$_\odot$ and  velocity v$_{out}^{comb}$=1280$\pm$480~km~s$^{-1}$, we estimate a global mass outflow rate of  $\dot{M}^{comb}_{[CO]}$ $\sim$139~M$_\odot$~yr$^{-1}$ and  a momentum boost  of $\dot{P}^{comb}_{[CO]}$/$\dot{P}_{[X]}$$\sim$30.

 Recent hydrochemical simulations coupled with an analytical model  (Richings \& Faucher-Gigu\`ere 2018a, 2018b) show that the gas swept-up by an AGN shocked wind is able to cool  and produce molecules within $\sim$1~Myr, and as a consequence, to give rise to powerful molecular outflow rates up  to $\sim$140M$_\odot$~yr$^{-1}$, strikingly similar to what observed in IRAS17.
Considering for example the AGN luminosity of $\sim$ 5$\times$10$^{44}$~erg~s$^{-1}$ and  the mass outflow rates  of the various components reported  in Table~\ref{tab:noema} for IRAS17, the analytic model (see Fig.10 in Richings \& Faucher-Gigu\`ere 2018b) is able to reproduce the behaviour of the NOEMA outflow, with the exception of the outflow velocities that are instead predicted to be much lower than the values measured in NOEMA spectra. As reported by these authors, the generally low value of the predicted outflow velocities ($\le$ 200 km~s$^{-1}$) compared to observations,  is likely due to the assumption of a uniform ISM in their modeling.

In IRAS17, the five NOEMA components with  their extremely high velocities reported in Table~\ref{tab:noema} can be interpreted as if they are steaming from the same outflowing gas that ``pierce"  through an inhomogeneous ambient medium giving rise to  apparently distinct molecular outflows. 
Given its high outflow velocity, it is likely that the CO gas seen by NOEMA is escaping mostly in a perpendicular direction with respect to the much denser molecular disc, along the path of least resistance, a scenario  remarkably consistent with what was postulated by Faucher-Giguere \& Quataert 2012 in their Fig.3.
Therefore,  the detection of the blue and red part of the ``molecular cone" can be regarded  as a natural consequence of the effect of the shocked wind bubble expanding normal to the galactic disc in a bipolar geometry, as envisaged e.g. by Faucher-Gigu\`ere \& Quataert 2012.  

\subsubsection{Witnessing the birth of  an inflating bubble?}
\label{sec:bubble}
We conclude this discussion by proposing a purely speculative yet plausible scenario that may provide further ideas to interpret current and future observations of AGN multi-phase outflows. 
According to the aforementioned model, as the shock keeps expanding beyond pc-scales, a bubble of tenuous hot plasma emitting in the X-rays is ``inflated" by energy deposition of the nuclear outflow.  
The signature of this extended X-ray emission provides the most compelling evidence of the interaction of the AGN outflow with the surrounding ISM (see Nims, Quataert \& Faucher-Gigu\`ere, 2015) yet it has been rarely detected  in external active galaxies (e.g. Greene et al. 2014, Fischer et al. 2019). However,  observing its most immediate fingerprint is actually readily accessible in the centre of our own Galaxy: indeed, the  {\it e-ROSITA} X-ray telescope has  observed two large scale bubbles of hot  gas emanating from the Galactic Centre through the Milky Way Halo (the so-called ``e-ROSITA bubbles", Predehl et al. 2020). 
Although alternative hypotheses have been proposed, these soft X-ray structures are commonly  interpreted as the echo of  nuclear past activity or strong star formation in the Galaxy (Yang et al. 2022).
The total extension of these almost bipolar X-ray bubbles is about 14~kpc and their partial overlapping with the {\it Fermi} bubbles discovered at energies higher than $\sim$1 ~GeV  (Su et al. 2010), has pointed to a common origin rooting in the past activity of the now dormant super massive black hole. 

In their analytical model of a quasar driven wind, several authors (e.g. Faucher-Gigu\`ere \& Quataert, 2012; Zubovas \& King 2014) present exhaustive arguments for these bipolar structures being produced by an expanding shocked outflow initiated by a fast X-ray wind in the past. Further theoretical arguments for energy injection by  supermassive black hole activity in Milky Way-like galaxies have been recently reported in the TNG50 cosmological simulations (Pillepich et al. 2021), which show that such expanding X-ray bubbles should be detectable in the local Universe with {\it e-ROSITA}.

 Thus, we can assume that  an X-ray UFO is a viable mechanism to inflate a high-energy bubble of hot gas in AGN. From an observational point of view, though, demonstrating the link between the X-ray UFO and the galaxy scale bubble  has not been  obvious because the detection of such  high-energy extended and tenuous structure is highly challenging in AGN, which are commonly dominated by the nuclear X-ray emission (see review by Fabbiano \& Elvis 2022).

   In this regard, recent results by the {\it Fermi-LAT} Collaboration have possibly provided evidence for the signature of the high-energy bubble in external galaxies where black hole activity and outflows are both traceable. A systematic search for Gamma-ray emission from a sample of 35 AGN selected to have an X-ray UFO  (see Ajello et al. 2021 for details) yields a positive Gamma-ray detection in their stacked signal of {\it Fermi-LAT} data, consistent with the expected UFO signature in this band.  On the contrary, the same analysis applied to a sample of AGN where the X-ray UFO is not present did not provide any  {\it Fermi-LAT} detection. The sources in both samples are Seyfert Galaxies with no individual detection of Gamma-ray emission from other processes.  In analogy to the mechanism that produced the {\it Fermi} bubbles, this result was interpreted by Ajello et al. (2021) as the signature of extended Gamma-ray emission produced via interaction of the X-ray UFO with its host galaxy.

When turning our eyes to IRAS17,  admittedly, current {\it Chandra} imaging is far from providing robust evidence for such X-ray bubble of gas, but neither could we demonstrate that the X-ray source is point-like  (see the detailed discussion in Section~\ref{subsec:xray}).  This indetermination leaves  more than an open path for the possibility that X-ray extended emission is present.  According to the theoretical expectations  above reported and inspired by the consolidated properties of the energy-conserving wind, we speculate that the putative diffuse X-ray emission could be the consequence of the shocked UFO that has started to inflate an {\it e-ROSITA/Fermi} bubble-like structure in this galaxy.

If confirmed, the presence of extended X-ray emission would provide the much needed missing probe that energy is effectively transported from the nuclear wind to the large scale phases of the outflow. 
Future  Chandra observations in a different instrumental configuration may corroborate the shocked outflow interpretation here proposed. 
\begin{table*}
  \centering
  \caption{Properties of the molecular outflows observed by NOEMA in IRAS17 estimated assuming a conversion factor $\alpha_{CO}$=0.8 M$_{\odot}$ (K km s$^{-1}$ pc$^2)^{-1}$.  
  A radius R= 2.8$\pm$0.3 kpc is assumed for estimating the combined mass outflow rate and momentum flux, everything else in calculated as per Table~\ref{tab:noema}.} 
  \begin{tabular}{lccccc}
    \hline \hline
    Outflow      &   $v_{peak}$         &       $M_{H_2}$            &    $\dot{M}_{H_2}$  & $\dot{P}_{[CO]}$ &  $\frac{\dot{P}_{[CO]}}{\dot{P}_{rad}}$ \\
     Component                       & ($km\,s^{-1}$)        & ($10^7\: M_\odot$)     &   ($M_\odot\,yr^{-1}$)  & ($10^{35}$  cm g s$^{-1}$)   & - \\ \hline
     {\bf B-W1}    &     $-1800\pm 35$   &   $5.3\pm 1.9$           &     $97\pm 26$      &  9.97$\pm$2.83     &  58$\pm$17   \\ %0.10+0.04
     {\bf B-W2}    &     $-610\pm 19$     &   $4.2\pm 1.3$           &     $25\pm 6$        &   0.87$\pm$0.22   &  5.11$\pm$1.4 \\ %0.07+0.03
     {\bf B-E}   &    $-1000\pm 10$    &   $2.04\pm 0.7$         &     $19\pm 5$        &    1.1$\pm$0.3   & 6.5$\pm$1.7 \\ %0.03+0.02
           \hline
     % {\bf  Total Blue}     &   1136$\pm$500 &     $11.5\pm3.9$   &     133[55--254]   &  8.6[2--23.6]     &  50 [11.7--138.8]     \\ 
     %  {\bf (W1+W2+E)}  \\
   % \hline\hline
      {\bf R-W} &  $1870\pm 20$      &   $1.36\pm 0.56$   &     $23\pm 7$   &  2.4$\pm$0.8   & $14^{+5}_{-4}$ \\
   {\bf  R-E}    &  $1100\pm 30$       &    $2.9\pm 0.9$      &     47$\pm7$    & 2.9$\pm$0.5   &   17$\pm$3 \\
   \hline
 %   {\bf  Total Red}     &   1485$\pm$380 &     $4.26\pm1.46$   &     71.6[41--104]   &  6[2.6--11]     &  35 [15--65]     \\ 
   %    {\bf (Red W+ Red E)}   \\
     \hline\hline
        {\bf Total } &   $1280\pm480$  &    15.8$\pm$5.4   &   220 [101 -- 367]   &   16 [4.6 -- 36.8] &    94 [27--216]  \\ 
   %   {\color{blue}   {\bf  \scriptsize{B\_(W1+W2+E)+R\_(W+E)}}} &  &   &  \\ 
                {\bf (All components) } &  &   &  \\ 
     \hline\hline
  \end{tabular}
  \label{tab:outflow_alpha0.8}
   \end{table*}

\subsection{Comparison of IRAS17 to other sources in literature}
\label{subsec:comparison}

In this section we discuss the results of the NOEMA campaign for IRAS17 within the context of the recent literature results on other sources. 
To this end, we have retrieved the observational facts reported by Marasco et al. 2020 in their Section 5.3 and Table B.1, where the compilation of the 8 sources\footnote{For completeness: we note that the sample presented in Marasco et al. 2020 is made by 10 sources including the two QSOs studied thereby, which present an outflow of ionized gas in optical data. We decided to carry out the comparison with IRAS17 using only data of neutral and molecular gas, therefore the sample used here is made by a total of 8 sources.} characterized by a solid detection of a nuclear X-ray wind and a galaxy-scale molecular/atomic outflow is presented. 
The properties of these sources in Marasco et al. 2020 were extracted from the literature and recalculated according to the same assumptions.  Analogously, the energetics of the UFOs and of the molecular outflows in these 8 sources  were recalculated so as to be comparable to those obtained in this work for IRAS17. Momentum rates of UFO were calculated following Equation 8 in Marasco et al. 2020. Since this compilation includes mostly powerful quasars and ULIRGS (Ultra Luminous Infrared Galaxies), their molecular outflows properties were estimated  assuming a luminosity-mass conversion factor of $\alpha_{CO}$=0.8 M$_{\odot}$ (K km s$^{-1}$ pc$^2)^{-1}$, which is  more typical of systems like submillimeter galaxies and starburst  (see Bolatto et al. 2013 for a thorough discussion on the $\alpha_{CO}$ conversion factor). A full description  of literature information on these 8 sources is reported in Marasco et al. 2020 (see their Appendix B), therefore the interested reader is deferred to this publication for more details. 
As to IRAS17, Table~\ref{tab:outflow_alpha0.8} reports the NOEMA outflow  properties rescaled to the mass obtained with the conversion factor $\alpha_{CO}$=0.8 M$_{\odot}$ (K km s$^{-1}$ pc$^2)^{-1}$. 
Then, the energetics estimated for the combined outflow in IRAS17 were plotted along with those of the other 7 sources to reproduce the diagram in Fig.~\ref{fig:ppunto_noema}.  
\begin{figure}
 \includegraphics[width=1.\columnwidth]{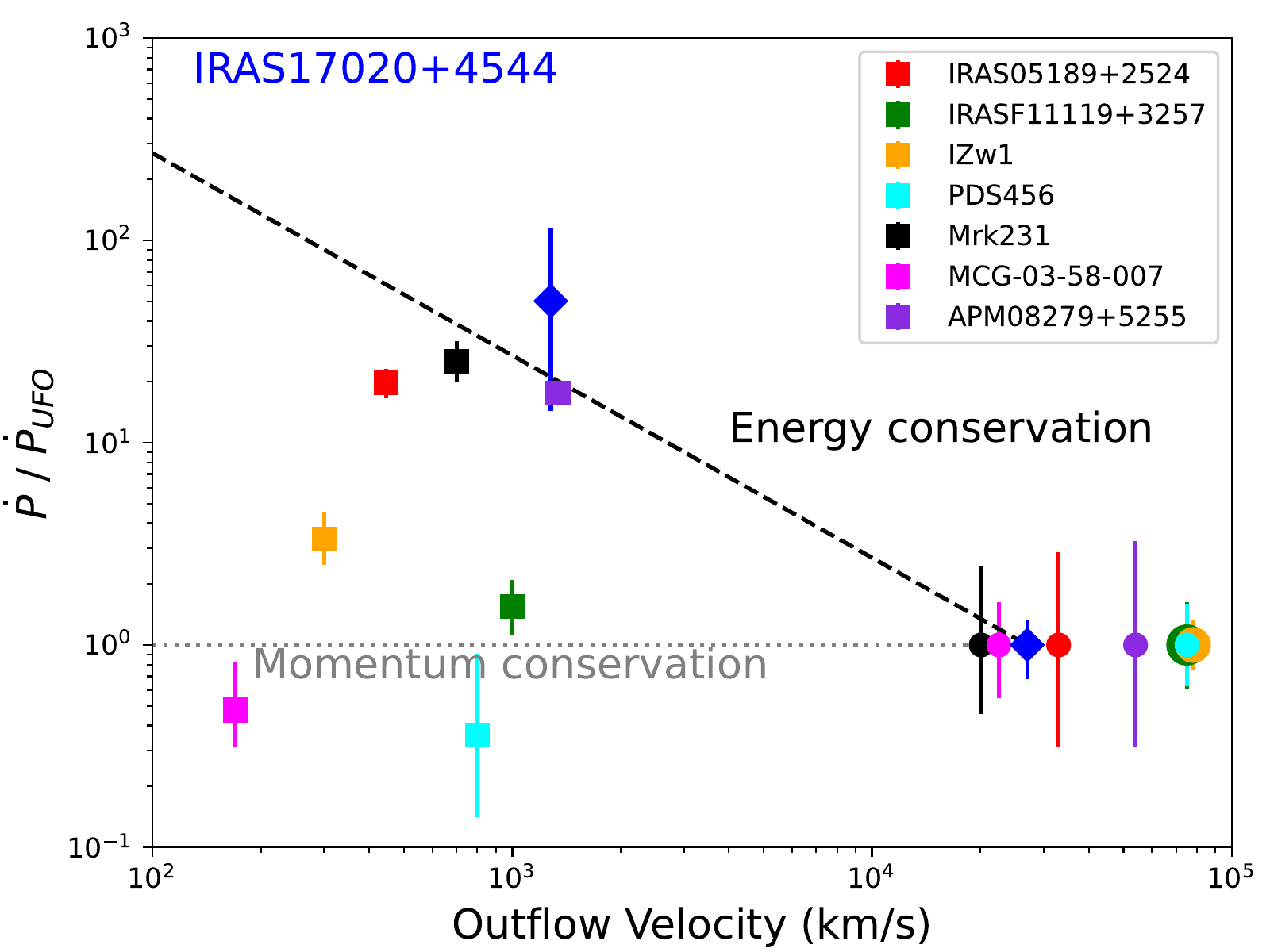}
 \caption{This plot shows the momenta of the molecular (squared marks) and X-ray (rounded marks) phases of the wind, rescaled to  each source's  X-ray momentum rate, as a function of the outflow velocities for sources where this relation has been observed. Blue diamonds mark 
  the combined molecular outflow observed by NOEMA in IRAS17 (Table \ref{tab:outflow_alpha0.8}) and the X-ray UFO  from Longinotti et al. 2015.
 The data for the other 7 sources are extracted  from Marasco et al. 2020 (see details in Section~\ref{subsec:comparison}).   
 The black dashed line marks the prediction for an energy conserving outflow i.e.  
$\dot{P}_{[CO]}$/$\dot{P}_{[X]}$=v$_{out\_X}$/v$_{out\_CO}$ in IRAS17. The grey dotted line marks the prediction for momentum-conserving outflows. 
}
\label{fig:ppunto_allsources}
 \end{figure}

 Figure~\ref{fig:ppunto_allsources} shows the energy vs momentum-conservation properties of the 8 AGN above described\footnote{For plotting purpose, the X-ray UFO velocity for PDS~456 was artificially shifted to a slightly lower velocity in order to not overlap with the points marking the two other sources with X-ray winds at 0.25{\it c}.}. 
 This plot seems to confirm an apparent dichotomy in the behaviour of the galaxy-scale outflowing gas:  IRAS17, Mrk231, IRAS05189+2524 and MCG-03-58-007 (all of a fairly diverse nature in terms of  bolometric luminosity, AGN type, BH mass) are consistent with energy-conservation, whereas the outflows in the rest of the sample  indicate a lower momentum boost. As already pointed out and discussed by Marasco et al. in their analysis,
 the cause of this dichotomy may be related to the fact that the relation plotted in  Figure~\ref{fig:ppunto_allsources} does not take into account that the galaxy-scale and nuclear outflows are likely subject to a different AGN radiation force.  These two phenomena act respectively at kpc and pc scales, implying that the bolometric luminosity of the AGN experienced by the molecular outflow and by the UFO may have varied over the respective flow timescales, an hypothesis consistent with the different outflow phases being powered by multiple episodes of AGN activity (see Zubovas et al. 2022).

\section{Summary}
\label{sec:summary}

This paper presents an analysis of the outflowing molecular gas traced by the CO(1-0) emission line  detected in the NOEMA data of  the Narrow Line Seyfert 1 Galaxy IRAS17020+4544.  Results from ancillary {\it Chandra} X-ray and e-MERLIN radio data are included in our discussion. 
The main conclusions can be summarized as follows: 

\begin{enumerate}
\item The molecular outflowing gas is  resolved in five components  distributed on the West  and East side of the active nucleus consistent with a biconical outflow with the approaching and receding side of the cone extending on a scale of at least 3~kpc from the nucleus. 
 The total mass in the combined outflow is M$_{CO}^{comb}$=(10$\pm$3.5)$\times$10$^7$M$_\odot$ and the average velocity is estimated to v$_{out}^{comb}$=1280$\pm$480~km~s$^{-1}$, although the highest velocities measured in two of the five components reach 1800-1900~km~s$^{-1}$.

\item The mass outflow rate estimated for the combined molecular outflow of the order of  $\dot{M}_{H_2}$=~139$\pm$20$~M_\odot$~yr$^{-1}$  coupled with the previous estimate of the X-ray UFO allow us to compare the momentum rates of the two outflow phases providing a definitive confirmation of the energy-conservation regime for   this powerful galaxy-scale outflow. The boost received by the momentum rate of the molecular outflow ($\sim$30) is interpreted as the manifestation of an efficient mechanism of the energy being transferred outward from the nuclear X-ray wind to the host galaxy. Evidence for such powerful outflow activity  provides a strong indication  that IRAS17 is undergoing feedback processes.

\item  The possible existence of extended X-ray emission hinted by {\it Chandra}  imaging and the presence of elongated radio emission on sub-kpc scale along with the NOEMA results and the previously reported X-ray UFO, are highly suggestive of a shock wind that is interacting with the ISM while expanding  within the central kpc of the galaxy, as postulated by several theoretical models (e.g. King (2010), Faucher-Gigu\`ere \& Quataert (2012)).

\end{enumerate}

\section*{Acknowledgements}
% Entry for the table of contents, for this guide only
\addcontentsline{toc}{section}{Acknowledgements}
 We are grateful to the anonymous referee for their careful review of the manuscript that significantly improved our publication.
This work is based on observations carried out under project number W17CR with the IRAM NOEMA Interferometer. IRAM is supported by INSU/CNRS (France), MPG (Germany) and IGN (Spain). This research has made use of data obtained from the Chandra Data Archive and the Chandra Source Catalog, and software provided by the Chandra X-ray Center (CXC) in the application package CIAO. 
The author is grateful to the Chandra-CXC Helpdesk and to H. Marshall and D. Huenemoerder for their support on the treatment of the HRC-LETG data.
A.L.L. and Q.S. acknowledge support from CONACyT grant CB-2016-01-286316.
Y.K. acknowledges support from  DGAPA-PAPIIT grant IN106518. 
C.F. acknowledge support from the PRIN MIUR project ``Black hole winds and the baryon life cycle of galaxies: the stone-guest at the galaxy evolution supper", contract 2017PH3WAT. 
 V.M.P.A. and V.C. acknowledge support from CONACyT research grants 280789 and 320987.
A.L.L. acknowledges the staff of the European Space Astronomy Centre (ESAC, Madrid) for hosting her visit during which this work was partly finalized. Financial support is acknowledged from ESA through the Science Faculty - Funding reference ESA-SCI-SC-LE-123,  and from project PID2019-107408GB-C41 by the Spanish Ministry of Science and Innovation/State Agency of Research MCIN/AEI/ 10.13039/501100011033.

\section*{Data availability}
The data underlying this article will be shared on reasonable request to the corresponding author.

%%%%%%%%%%%%%%%%%%%%%%%%%%%%%%%%%%%%%%%%%%%%%%%%%%

%%%%%%%%%%%%%%%%%%%% REFERENCES %%%%%%%%%%%%%%%%%%

% The best way to enter references is to use BibTeX:

%\bibliographystyle{mnras}
%\bibliography{example} % if your bibtex file is called example.bib

% Alternatively you could enter them by hand, like this:

\appendix
 \section{Ancillary data}
\label{appendix}

\subsection{The {\it Chandra} X-ray emission: a possible extended component?}
\label{subsec:xray}
 \begin{figure}
 \includegraphics[width=1.1\columnwidth]{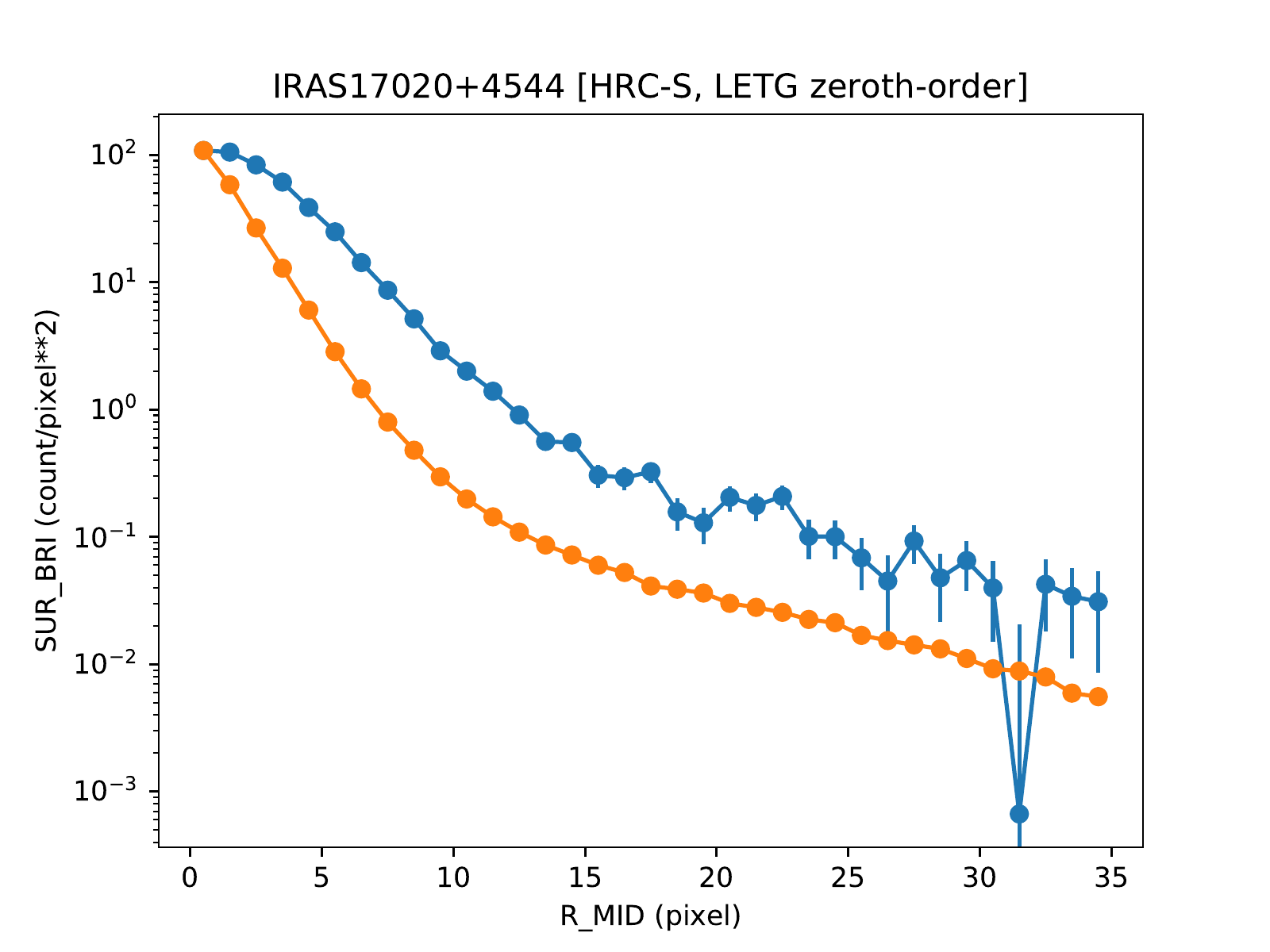}
  \includegraphics[width=1.1\columnwidth]{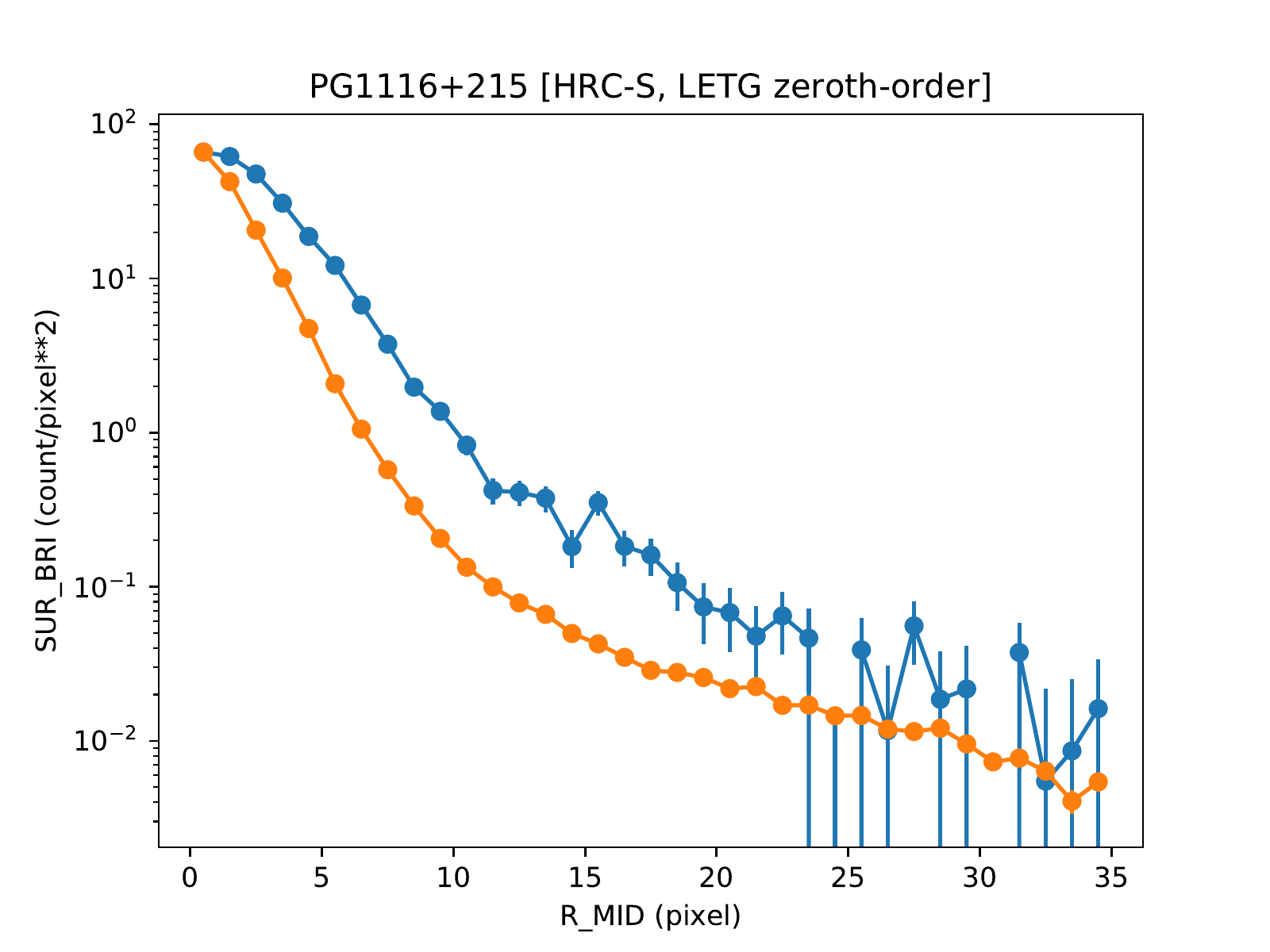}
   \includegraphics[width=1.1\columnwidth]{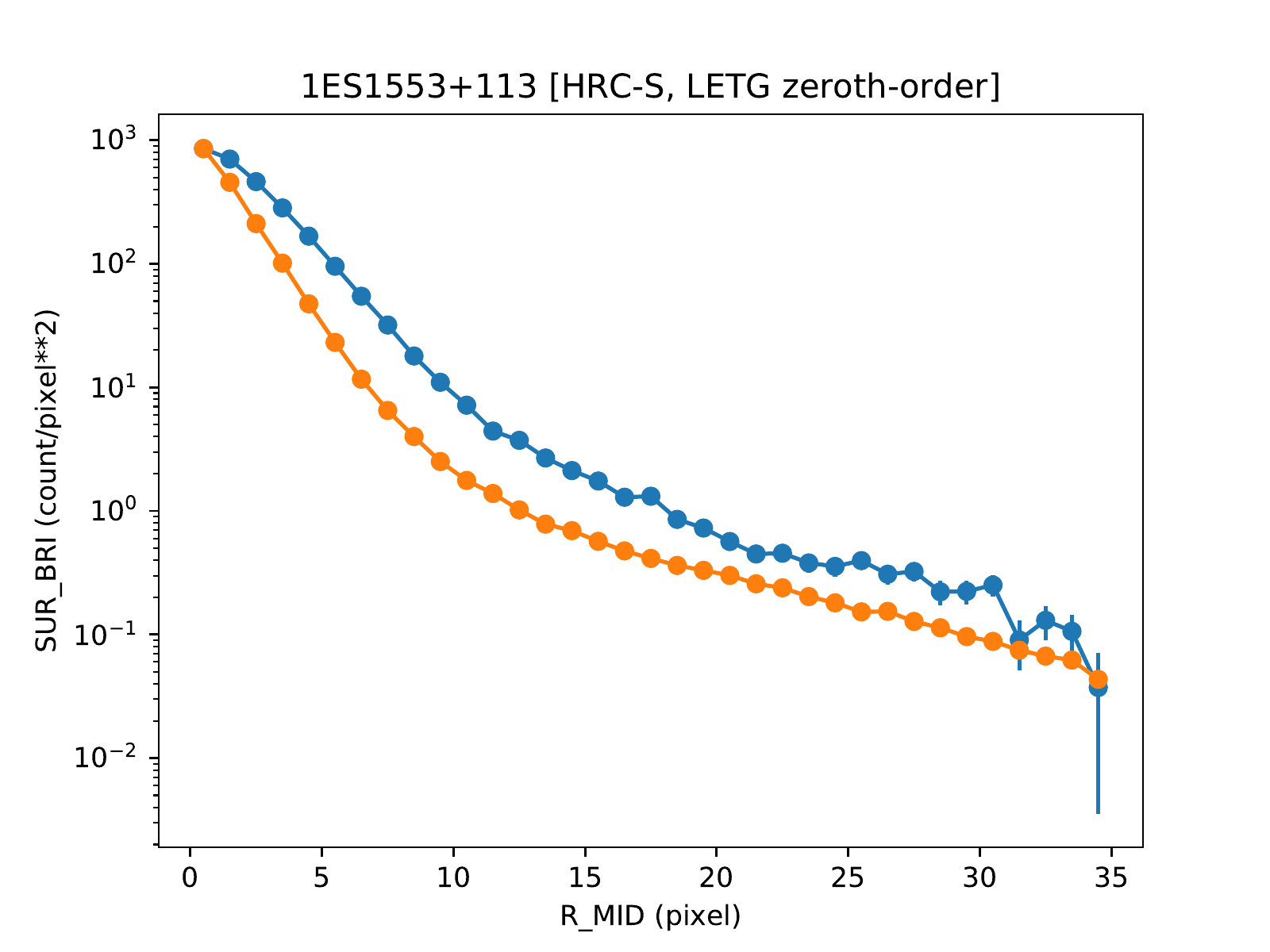}
  \caption{Comparison of the radial brightness profile of the source (blue) and PSF (orange) obtained by ray-tracing with ChaRT for IRAS17 and for the other two comparison sources: PG1116+215 and  1ES1553+113 }
 \label{fig:rad_prof}
\end{figure}
{\it Chandra} observed IRAS17 in 9 visits distributed from November 2016 to February 2017 and for a total exposure of 250~ks (Sequence Number 703217). 
As mentioned in Section \ref{subsec:Chandra},  all observations were accomplished with the HRC-S camera as readout detector for the LETG grating. The data reduction was carried out using the script  {\tt chandra\_repro} with calibration version CALDB 4.8.4.1,  which are provided within the software CIAO 4.11 (Fruscione et al. 2006). The output of this script is a new level=2 event file, that was then input into the CIAO tool  {\tt dmcopy} to extract the zeroth order image of the source onto the HRC detector. 

Since the focus of this analysis is the imaging information, details on the data reduction and analysis of the spectral products are not reported. 
Preliminary results on the spectroscopic information extracted by these data are available in Longinotti (2020).
 The zeroth order images were produced for each of the 9 Chandra segments. They were extracted in the 0.1-10 keV energy band and with a spatial resolution of 0.4~arcsec, which correspond to the nominal spectral band and resolution of the HRC detector.

As can be seen in Fig.~\ref{fig:Chandra}, the X-ray image os IRAS17  tentatively shows a slight elongation in the NE direction.  A visual examination of  the zeroth order images provided by the {\it TGCat} database  (Huenemoerder et al. 2011) of several other sources (mostly AGN and stars) observed in the same instrumental configuration (HRC-S+LETG) revealed to us that IRAS17 is basically the only source showing such apparent asymmetry. These images are available in the conference proceeding Longinotti (2022). Giving the prior evidence for nuclear and galaxy scale outflows occurring in this source, the existence of diffuse X-ray emission was definitely an intriguing possibility to investigate. 

In an attempt to detect possible extended emission in the {\it Chandra} data we then performed a study of the brightness radial profile of the source. This profile was extracted from a series of 35 concentric annuli centred on the galaxy nucleus. Then,  it was  compared to the brightness profile of the PSF produced by the Chandra Ray Tracer ({\it ChaRT}) (Carter et al. 2003)  and {\it MARX} (Davis et al. 2012)  tools assuming the well characterized 0.3-10~keV spectral shape from the archival EPIC {\it XMM-Newton} observations  described in Longinotti et al. 2015. This assumption is justified by the lack of flux and spectral variability between the {\it Chandra} and {\it XMM-Newton} observations (see Section 3 in Longinotti 2020). This test was initially carried out on the longest {\it Chandra} segment  (OBS ID 18152 for an exposure of 55~ks, see top panel in  Fig.\ref{fig:rad_prof}), and then repeated on the other 8 segments. The source radial profile extracted from all 9 segments shows indeed tantalizing evidence for extension out to 20 px from the centre. 
 
  However, it is well known that the presence of the grating coarse support structure produces a spiked pattern in the zeroth order image, therefore preventing  a  ``clean" analysis of the spatial distribution of the counts also in the simulated PSF.  To test for an instrumental effect responsible for the apparent extension of the source in the radial profile, we run the same procedure in two point-like sources (the Quasar PG1116+215 and the Blazar 1ES1553+113) observed with the same instrumental configuration (HRC-S +LETG) and with an X-ray spectrum similar to IRAS17. In both sources this  test shows the same apparent extension in the source brightness radial profile when compared to the simulated  HRC PSF (see central and bottom panels in Fig.\ref{fig:rad_prof}), suggesting that an instrumental effect is likely causing the apparent source extension. 
  
This test shows that  based on the brightness radial profile extracted from the currently available image, we cannot tell whether IRAS17 presents extended X-ray emission,  but neither can we establish if the X-ray emission of this source is point-like.  This led us to conclude that the presence of diffuse X-ray emission cannot be formally excluded with the present data.

\subsection{e-MERLIN  observations}
\label{subsec:radio}

\begin{figure}
	\includegraphics[width=8cm]{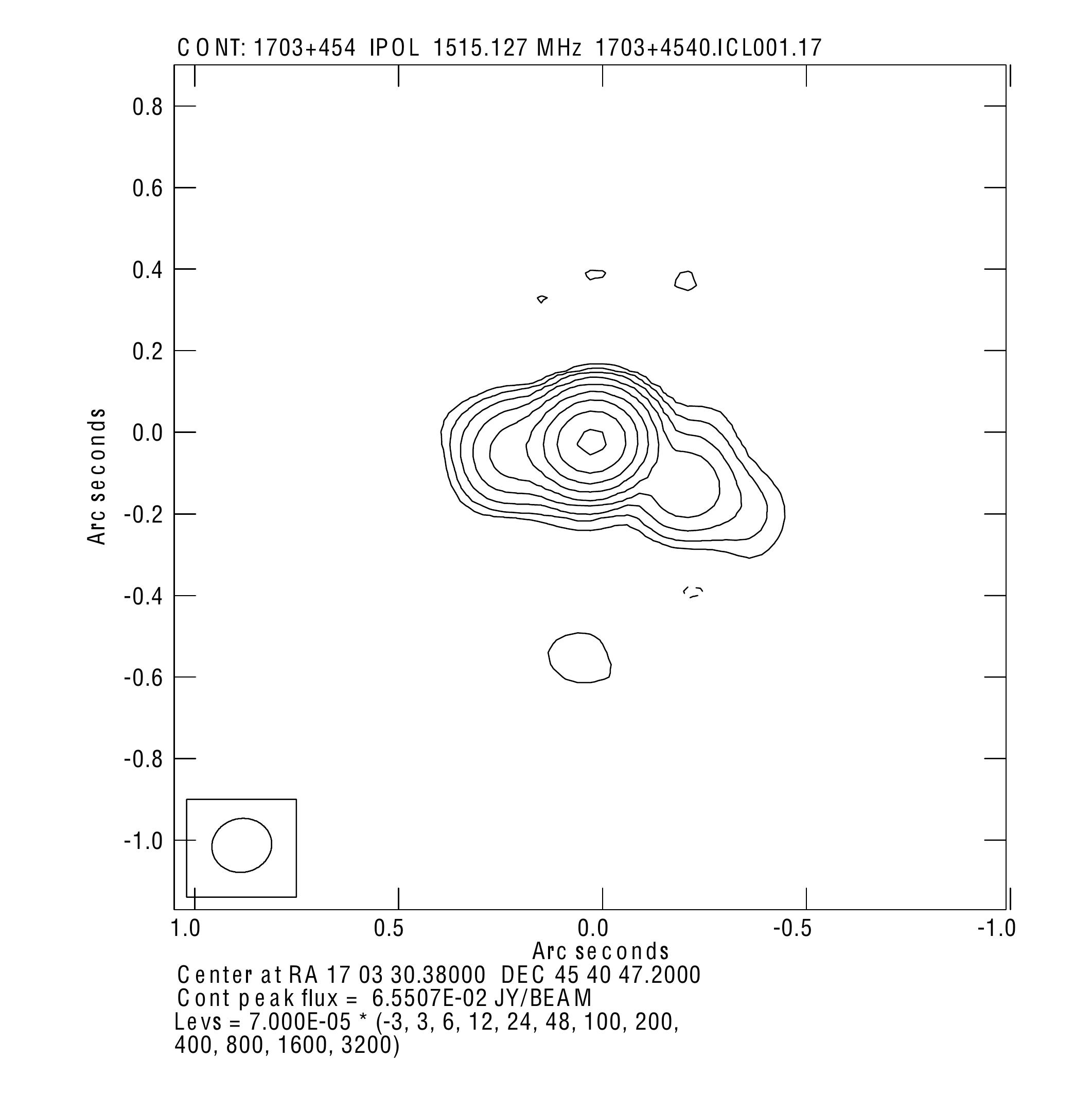}
			\caption{Image of the source  at 1.5 GHz from the e-MERLIN campaign showing a dominant central core and an apparent elongation on both sides, more pronounced towards south-west.  This image is the same as in Fig.~\ref{fig:Chandra}, only with a slightly different choice of contour levels. }
		\label{fig:emerlin}
	\end{figure}
IRAS17 has been the target of an extensive radio campaign that is still ongoing and that will be presented in a forthcoming publication (Stanghellini et al. in prep).
Here we summarize the main findings of the e-MERLIN data analysis that were used to draw the contours of the radio source at $\sim$ 1arcsec scale (see Fig.\ref{fig:Chandra}). 
e-MERLIN observations at 1.51 GHz were carried out on 14 February and 12/13 March 2020. 
The summary of the observations is given in Table~\ref{tab:obs_merlin}.
Data have been provided already calibrated with the standard pipeline. Imaging and self-calibration have been done with Difmap and AIPS.

Fig.~\ref{fig:emerlin} shows that the radio source at this scale is composed by a dominant central core and two secondary components located East and SW of the core. 
 
A simple fit with 3 components (modelfit in difmap) has been done, with an unresolved component (delta function) for the core and 2 circular gaussian components for the secondary components. Considering the wide frequency bandwidth and the partial blending of the components, the flux densities do not have a high accuracy (10\% to be conservative). Flux densities of the components are given in Table~\ref{tab:merlin_fluxes}.
By comparison, the NVSS and FIRST surveys provide 1.4 GHz flux densities of 121 and 118 mJy, respectively (Giroletti et al. 2017).

 \begin{table}
	%\begin{center}
		\caption{e-MERLIN observation log}
		\label{tab:obs_merlin}
		\begin{tabular}{c|c|c|c} 
		 \hline\hline
			 date & $\nu$&$\Delta\nu$& Exposure \\
			      &     GHz  & GHz&   \\
			\hline\hline	
			14-Feb-2020&         1.51   & 0.5&17.5h\\
			18/19-Feb-2020&  1.51& 0.5& 9.5h \\
			12/13-Mar-2020& 1.51&0.5&18.5h\\
			\hline
		\end{tabular}
%	\end{center}
\end{table}

\begin{table}
%	\begin{center}
		\caption{e-MERLIN flux densities and size}
		\label{tab:merlin_fluxes}
		\begin{tabular}{c|c|c} 
		\hline \hline
			Component  & $S_\mathrm{1.51\,GHz}$    & size \\
			 - & [mJy]   &[arcsec] \\
			\hline\hline
			\textbf{core} & 63.4 & -\\
			\textbf{East} & 8.7  & 0.17\\
			\textbf{SW}  & 6.2   & 0.15 \\
			\hline
		\end{tabular}
%	\end{center}
\end{table}

%------------------------------------------------

%%%%%%%%%%%%%%%%%%%%%%%%%%%%%%%%%%%%%%%%%%%%%%%%%%

%%%%%%%%%%%%%%%%% APPENDICES %%%%%%%%%%%%%%%%%%%%%

\end{document}